\documentclass[10pt,a4paper]{article}

\usepackage[T1]{fontenc}
\usepackage[utf8]{inputenc}
\usepackage{amssymb,amsmath,amsthm,amsfonts}    
\usepackage{bm}                                 
\usepackage{mathtools}                          
\usepackage{stmaryrd}                           
\usepackage{breqn}                              
\usepackage[textfont=footnotesize,labelfont={footnotesize,bf}]{caption}
\usepackage{subcaption}
\usepackage{booktabs}
\usepackage[inline]{enumitem}
\usepackage{graphicx}
\usepackage[dvipsnames]{xcolor}
\usepackage[top=30mm,right=30mm,left=30mm,bottom=30mm]{geometry}
\usepackage[colorlinks=true,linkcolor=blue,urlcolor=blue]{hyperref}
\usepackage[affil-it,auth-sc]{authblk}
\usepackage[colorinlistoftodos,textwidth=25mm,textsize=footnotesize]{todonotes}
\usepackage{lipsum}                             
\usepackage[section]{placeins}
\usepackage[
	backend=biber,
	style=numeric,
	citestyle=numeric-comp,
	maxbibnames=99,
	minbibnames=99,
	maxcitenames=2,
	mincitenames=1,
	giveninits=true,
	sorting=none,
	sortlocale=auto,
	natbib=true,
	url=false,
	isbn=false,
	doi=true,
	eprint=true
]{biblatex}
\newcommand{\trans}[1]{#1^{\mathsf{T}}}

\newcommand{\nderiv}[3]{\frac{\partial^{\,#3} #1}{\partial #2^{\,#3}}}

\newcommand{\ba}{\bm a}
\newcommand{\bb}{\bm b}

\newcommand{\be}{\bm e}
\newcommand{\bef}{\bm f}

\newcommand{\bq}{\bm q}
\newcommand{\br}{\bm r}

\newcommand{\bt}{\bm t}
\newcommand{\bu}{\bm u}

\newcommand{\bx}{\bm x}

\newcommand{\bB}{\bm B}

\newcommand{\bE}{\bm E}

\newcommand{\bI}{\bm I}

\newcommand{\bK}{\bm K}

\newcommand{\bN}{\bm N}

\newcommand{\bZ}{\bm Z}

\newcommand{\fE}{\mathbb E}

\newcommand{\mC}{\mathcal C}

\newcommand{\mE}{\mathcal E}

\newcommand{\mW}{\mathcal W}

\graphicspath{{./}{./figures/}}                 

\renewcommand{\epsilon}{\varepsilon}

\title{Homogenization of architected materials incorporating shearable beams}

\author[1]{Matteo Franzoi}
\author[1]{Davide Bigoni\footnote{Corresponding author: e-mail: \href{mailto:name@unitn.it}{bigoni@ing.unitn.it}; phone: +39\,0461\,282507.}}
\author[1]{Andrea Piccolroaz}

\affil[1]{Instabilities Lab, University of Trento, Via Mesiano 77, 38123 Trento, Italy}

\date{\today}

\begin{document}

\maketitle

\begin{abstract}
\noindent
Two-dimensional architected materials are often realized as periodic grids of elastic beams. Conventional homogenization methods represent these structures as equivalent elastic solids but neglect shear deformation in the constituent beams. This article addresses this limitation by incorporating shear deformability through Timoshenko beam theory, enabling accurate modeling of stubby beams. 
Moreover, shearable beams with extreme mechanical characteristics can be obtained through the design of appropriate microstructures. Introducing shearable beams into the grid expands the design space, allowing, for instance, the control of the effective Poisson's ratio beyond the limits achievable with slender beams. 
\end{abstract}

\paragraph{Keywords}
Architected materials \textperiodcentered\
Timoshenko beam \textperiodcentered\
Homogenization \textperiodcentered\
Auxetic materials

\section{Introduction}
\label{sec:introduction}
Homogenization of composites is a key tool in the design of architected materials with predetermined mechanical properties. In this context, elasticity is particularly appealing because architected materials often must withstand cyclic loading and vibrations without undergoing permanent deformation. Although homogenization theory is well established within the framework of linear elasticity \cite{abdoul-anziz_homogenization_2018, castaneda_nonlinear_1997, elsayed_analysis_2010, hutchinson_structural_2006, milton_theory_2002, nemat-nasser_micromechanics_1993, willis_elasticity_1982, willis_mechanics_2002}, explicit closed-form solutions for equivalent materials are scarce, and several aspects still merit investigation. 
This article extends the results obtained for two-dimensional periodic grids of elastic beams introduced in \cite{bordiga_dynamics_2021, viviani_homogenization_2024} by incorporating the Timoshenko theory of shear-deformable beams. 

The interest in the modeling of grids composed of Timoshenko beams arises from three distinct sources.
\begin{enumerate}[label=(\roman*)]
    \item Domokos \cite{Domokos1993, Domokos1993109}, followed by Challamel and co-workers \cite{Kocsis, Kocsis2018, Lerbet20201}, and Paradiso et al.~\cite{paradiso}, have shown that homogenization of a discrete and periodic chain of rigid elements with lumped deformability leads to a continuous rod model that can be represented by either the unshearable Euler model or the Timoshenko model, depending on the mechanical characteristics of the junctions in the chain. In particular, the Timoshenko model is obtained with a chain such as that shown in Fig.~\ref{trave_a_taglio}, equipped with hinges (stiffness $k_r$), sliders (stiffness $k_t$), and axial springs (stiffness $k_a$). In the homogenized continuous model derived from the discrete structure, the bending stiffness \lq $EJ$' is given by $k_r a$, while the shear stiffness \lq $\kappa\, GA$' and axial stiffness \lq $EA$' are replaced by $k_t a$ and $k_a a$, respectively, where $a$ is the length of the unit cell. 
    %
    \begin{figure}[hbt!]
    \centering
    \begin{subfigure}[]{0.60\textwidth}
        \centering
        \includegraphics[width=\textwidth, keepaspectratio]{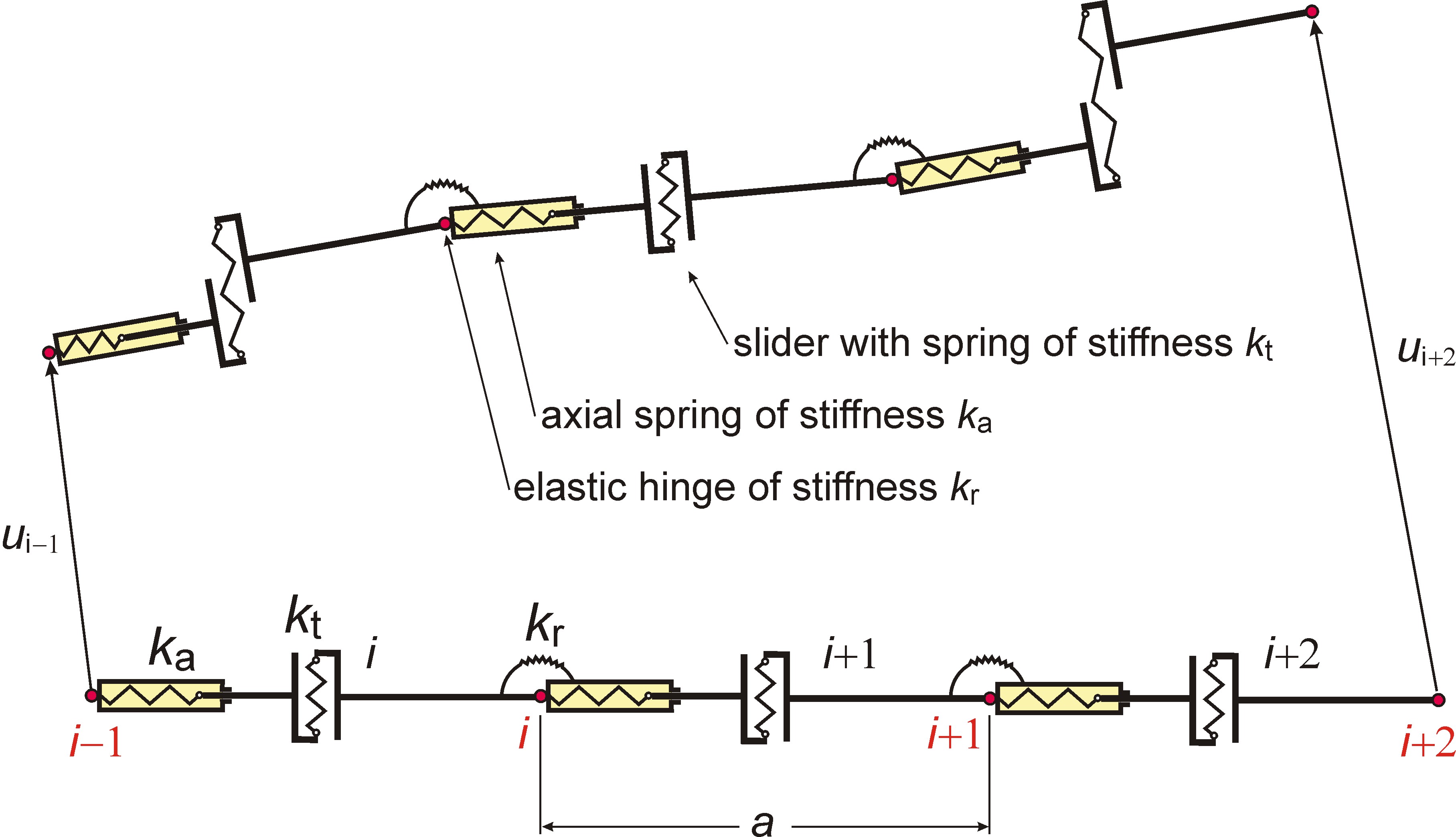}
    \end{subfigure}
    \caption{
    A microstructured chain model, represented from node $i-1$ to node $i+2$.
    The chain is constructed by periodically repeating a unit cell of width $a$; this cell comprises rigid bars connected by axial springs with stiffness $k_a$, elastic hinges with stiffness $k_r$, and elastic sliders (that permit only transverse relative displacement) with stiffness $k_t$.
    The $i$-th node undergoes a displacement $u_i$ and the overall behavior of the chain can be homogenized, resulting in the Timoshenko beam model, with bending, shear, and axial stiffnesses equal to $k_r a$, $k_t a$, and $k_a a$, respectively; these stiffnesses remain mutually independent.
    }
    \label{trave_a_taglio}
    \end{figure}
    %
    This result, detailed in Section~\ref{sec:decoupled_plane_stress}, shows that designing and manufacturing elastic beams in which {\it the three stiffnesses are mutually independent} is possible. This opens the possibility of creating slender beams whose shear compliance can range from zero to infinity, potentially dominating the axial and bending responses. In such a configuration, the beam slenderness is given by $\Lambda = l \sqrt{k_a/k_r}$ and may reach very low values, approaching zero. 
    \item Beyond the approach described above, another option exists. A grid of stubby elastic rods can be realized by superimposing beams on different planes so that they interact only through their end cross-sections. This configuration enables levels of stubbiness that cannot be achieved by simply perforating a plate. In particular, the slenderness can approach the theoretical minimum value of 2 (for a hexagonal grid). This arrangement yields a homogenized response with mechanical properties beyond the capabilities of grids of slender beams. This is explained in detail in Section~\ref{sec:decoupled_plane_stress} and illustrated in Fig.~\ref{fig:decoupled_model}.
    \item Finally, for grids of beams lying in the same plane (i.e., obtained by perforating a plate), there is a range of beam stubbiness for which the Euler model provides unrealistic values, while the Timoshenko model maintains a correct behavior for slenderness $\Lambda < 12$, as shown in Fig.~\ref{fig:triangular_lattice_plane_stress_20_cells}.
\end{enumerate}

\begin{figure}[hbt!]
    \centering
    \begin{subfigure}[]{0.96\textwidth}
        \centering
        \includegraphics[width=\textwidth, keepaspectratio]{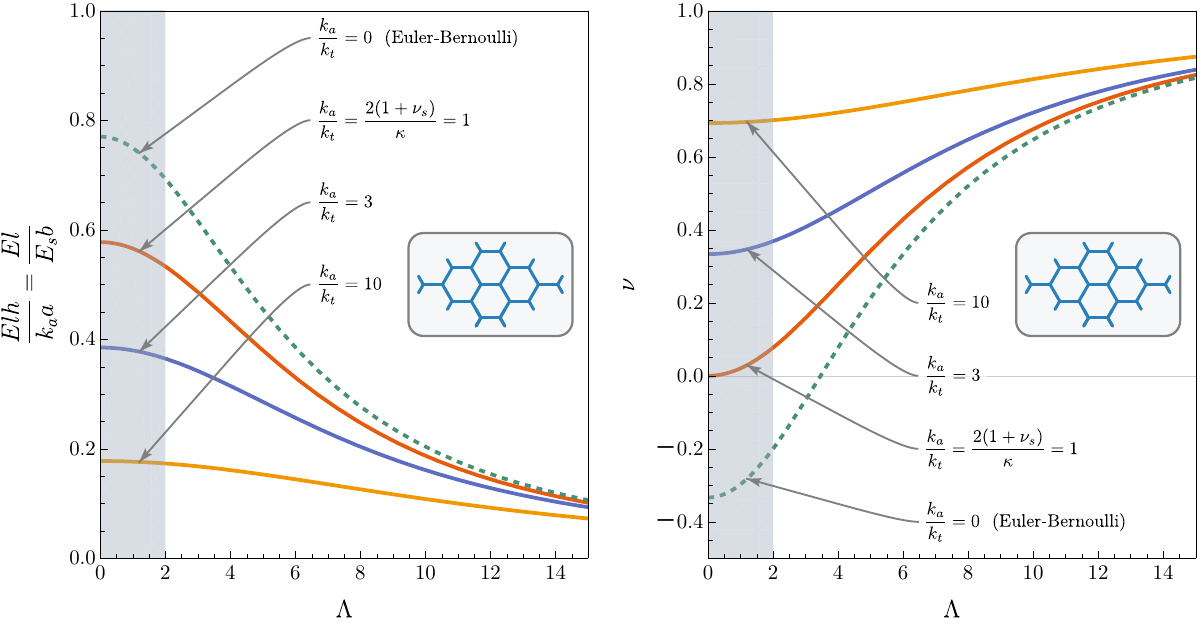}
    \end{subfigure}
    \caption{Effective elastic modulus $E$ and Poisson's ratio $\nu$ characterizing an elastic solid of out-of-plane thickness $h$, equivalent to a two-dimensional grid of beams arranged in a hexagonal lattice (inset). 
    Two models are considered for the constituent beams: (i) a homogenized discrete chain, characterized by the slenderness $\Lambda=l \sqrt{k_a/k_r}$, where $k_a$, $k_t$ and $k_r$ are the stiffnesses of the axial spring, slider spring, and elastic hinge, respectively, and $l$ is the total chain length, see Fig.~\ref{trave_a_taglio}; (ii) a continuous elastic beam of length $l$, cross-section area $A = b h$, slenderness $\Lambda=2\sqrt{3}\,l/b$, and shear correction factor $\kappa$, made of an isotropic elastic material with Young's modulus $E_s$ and Poisson's ratio $\nu_s$ (the grid is obtained by superimposing beams on different planes, as explained in Section~\ref{sec:decoupled_plane_stress}). 
    Different curves represent various ratios $k_a/k_t$ ($E_s A/(\kappa\, G_s A)$ for the continuous model), with dashed lines indicating the limiting case $k_t \to \infty$ (or $\kappa\, G_s A \to \infty$), which corresponds to the Euler–Bernoulli theory.
    }
    \label{fig:triangular_lattice_compare_material}
\end{figure}

Fig.~\ref{fig:triangular_lattice_compare_material} highlights the advantages of the Timoshenko beam model by plotting the effective elastic modulus $E$ (left) and Poisson's ratio $\nu$ (right) for an isotropic solid equivalent to a two-dimensional grid of elastic rods of length $l$ and slenderness $\Lambda$, arranged in a hexagonal geometry. The rods are either microstructured as in Fig.~\ref{trave_a_taglio} or elastic beams with Young's modulus $E_s$, Poisson's ratio $\nu_s$, and rectangular cross section $b \times h$. While microstructured beams can achieve any slenderness $\Lambda$, the method of superposition of elastic beams on different planes cannot permit slenderness $\Lambda$ smaller than 2 (region highlighted in gray in the figure). The various curves in the figure correspond to the Timoshenko beam model for different values of $k_a/k_t = 2(1+\nu_s)/\kappa$, whereas the case $k_a/k_t=0$ represents the Euler-Bernoulli model.

For the range of slenderness $\Lambda$ shown in the figure, the Euler–Bernoulli theory tends to overestimate the effective elastic modulus and may provide inaccurate values for the effective Poisson’s ratio. 
The Euler–Bernoulli model also predicts a negative effective Poisson’s ratio, suggesting auxetic behavior, whereas the correct values are positive. Moreover, in the Euler-Bernoulli model, the effective Poisson's ratio $\nu$ is unaffected by the constituent stiffness ratio $k_a/k_t$ (or $\nu_s$), whereas the Timoshenko theory shows a strong dependency on it. 
This example reveals a valuable design opportunity. Using microstructured or stubby beams with high shear deformability, the equivalent material can achieve high stiffness and Poisson's ratio values significantly different from those of slender-beam assemblies. 
The present article shows that proper design makes grids of microstructured or stubby beams feasible (Section \ref{sec:decoupled_plane_stress}) and that employing the Timoshenko model (Section \ref{sec:timoshenko_beam}) combined with homogenization (Section \ref{sec:homogenization}) leads to mechanical characteristics unachievable with assemblies of slender beams (Section \ref{sec:influence_shear}). These characteristics can involve all elastic parameters of the equivalent solid and highlight the role that the grid constituents play in its overall behavior as a solid.
It is noticed in closure that the approach to homogenization developed here is based on the assumption that the characteristic size of the microstructure is much smaller than the specimen size. No attempt is made to develop a higher-order approach, as in \cite{bacca}.

\section{Timoshenko model for grids of microstructured beams and for grids of stubby beams}
\label{sec:decoupled_plane_stress}
This section elaborates on points (i)–(iii) presented in the introduction. First, it briefly demonstrates how a discrete chain of rigid elements with lumped deformation sources can be converted into a continuous Timoshenko beam model in which axial, shear, and bending deformations are defined by independent parameters. Second, it explains, at least in principle, how to realize a junction between stubby beams working in different planes. Third, it derives the homogenization scheme for grids of Timoshenko shearable beams.

\subsection{Continualization of a microstructured chain into a beam model}
An axially and shear-deformable rod model is obtained from the discrete chain shown in Fig.~\ref{trave_a_taglio}. The elastic energy stored in a discrete chain of $n$ elements with lumped deformability, starting at node 0 and element 1, and ending at node $n$ and element $n$, is 
\begin{equation}
    \mE = 
    \frac{1}{2} \sum_{i=1}^{n} \Big[k_a \big(u_i-u_{i-1}\big)^2 + 
    k_t \big(v_i-v_{i-1}-a \theta_i\big)^2 \Big] + 
    \frac{k_r}{2} \sum_{i=1}^{n-1} \big(\theta_{i+1}-\theta_{i}\big)^2, 
\end{equation}
where $\theta_i$ is the rotation of the $i$--th element, and $u_i$ and $v_i$ are the horizontal and vertical components of the displacement of the $i$--th node. In the limit as $a \to 0$ and $n \to \infty$, the sums become integrals and the differences become derivatives of continuous functions of the axial variable $x$, yielding
\begin{equation}
\label{discretazzo}
    \mE = \frac{a}{2}\int_0^l \Big[
    k_a u^{\prime\,2} + 
    k_t \big(v' -\theta\big)^2 +
    k_r \theta^{\prime \,2}
    \Big] dx. 
\end{equation}
Eq.~\eqref{discretazzo} models the mechanics of a Timoshenko beam in which the bending stiffness $E_sJ$, shear stiffness $\kappa\, G_s A$, and axial stiffness $E_sA$ correspond to $k_r a$, $k_t a$, and $k_a a$, respectively. Importantly, the microstructured chain enables a beam whose bending, shear, and axial stiffnesses can be adjusted independently over the entire range from zero to infinity.

\subsection{A design strategy for grids with stubby beams}
\begin{figure}[hbt!]
    \centering
    \begin{subfigure}[]{0.96\textwidth}
        \centering
        \includegraphics[width=\textwidth, keepaspectratio]{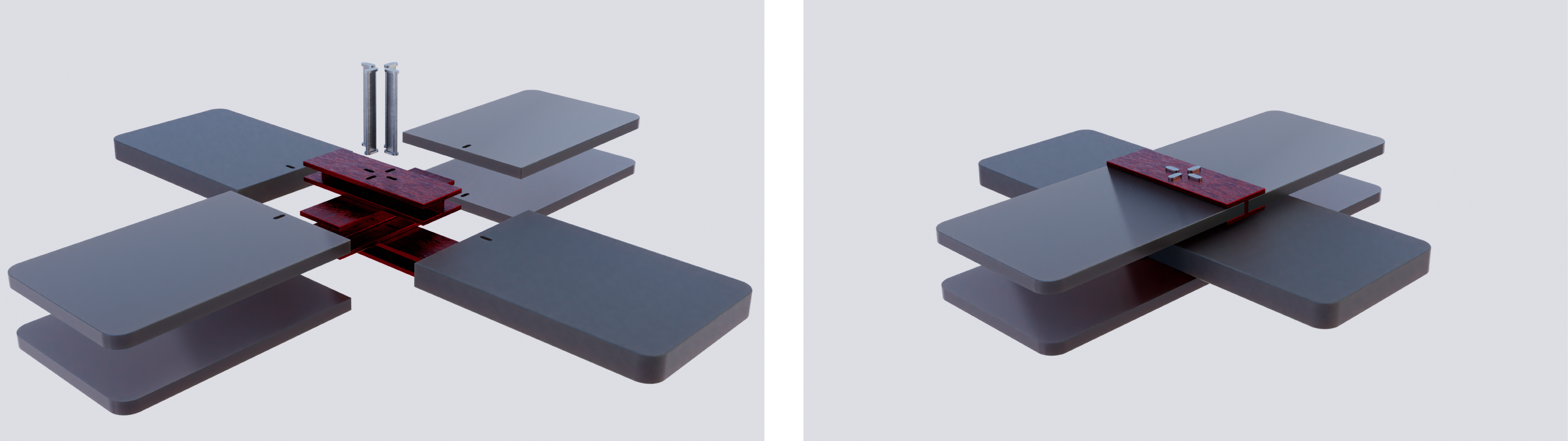}
    \end{subfigure}
    \caption{
    Concept model of a nodal connection where four stubby beams are jointed. The connection is designed for two orthogonal rods as a multi-layer rigid joint (colored in purple),  where the out-of-plane symmetry is preserved: (left) exploded view; (right) perspective view.
    }
    \label{fig:decoupled_model}
\end{figure}

Two-dimensional grids of beams can be realized by perforating a plate; however, this is not the only possible approach. In particular, Fig.~\ref{fig:decoupled_model} shows a proof-of-concept realization of a node connecting four stubby beams. The key idea for the practical realization of the junction is to use thin elements operating in different planes, but preserving the out-of-plane symmetry. 

The junction imposes full continuity of displacement between connected elements, which are reinforced with stiff coatings and linked by orthogonal \lq pivots' with rectangular cross-section to prevent relative rotation, in addition to relative displacement. In this arrangement, in-plane axial and shear forces, as well as bending moments, are fully transmitted through the node, 
while out-of-plane forces and moments do not develop, provided that the load is symmetric in the plane.
Such junctions must be rigid (made of a material stiffer than the beams) and allow the construction of grids of elastic stubby beams that are mechanically equivalent to a uniform material under plane stress conditions.
 Special junctions connecting elements located in different planes are required because, when a grid is cut directly from a plate, the spaces between beams shrink as the beams become stubby, invalidating the beam model; in this case, a model of a plate with holes becomes more appropriate, as shown in the insets of Fig.~\ref{fig:triangular_lattice_plane_stress_20_cells}. 

Figure \ref{fig:decoupled_model} illustrates a simple case in which the beams converging at the node have a rectangular cross-section, for which the shear correction factor is $5/6 \approx 0.833$. Under these conditions, there exists a lower limit for the slenderness, corresponding to the situation where parallel coplanar beams touch each other, thus occupying all the space between two consecutive lines. For a square grid, the rectangular cross-section assumption yields a minimum slenderness $\Lambda$ of $2\sqrt{3} \approx 3.46$ [see Eq.~\eqref{costantazze1}]. The minimum slenderness is equal to 6 for an equilateral triangular lattice and 2 for a hexagonal grid. Smaller values can be achieved by adopting different shapes of the cross-section. For example, a double-T section allows a slenderness of 2 for the square lattice; however, in this case, the shear correction factor drops to approximately $0.296$. For the sake of simplicity, this article considers only continuous elastic beams with rectangular cross-sections, although using microstructured beams would enable achieving arbitrarily small slenderness values. Therefore, results are presented starting from zero slenderness to provide a more comprehensive view of the situation.

\subsection{A validation of the homogenization scheme for grids of Timoshenko beams}
Although it is well known that the Timoshenko beam model can effectively predict the quasi-static and dynamic responses of stubby beams \cite{diaz-de-anda_experimental_2012,labuschagne_comparison_2009,mendez-sanchez_experimental_2005,renton_check_2001,stephen_check_2002}, a validation of the homogenization scheme proposed in the present article, specifically through the use of the Timoshenko model, has not been previously presented. Therefore, although not a primary focus of this study, finite element computations are included in Appendix~\ref{sec:appendix}, referring to an equilateral triangular grid of beams. By considering a finite yet sufficiently large grid of elastic rods --such that end effects can be neglected-- the equivalent solid obtained through homogenization is shown to accurately capture the overall behavior of the ensemble. At the same time, the limitations of the approach when a perforated plate is modeled as a beam network are also highlighted.

\section{The Timoshenko beam model}
\label{sec:timoshenko_beam}
The Timoshenko beam theory relaxes the Euler-Bernoulli assumption that the cross sections 
of a deflected beam remain perpendicular to its axis. In particular, for a rectilinear beam belonging to the $x$--axis (parallel and orthogonal to the unit vectors $\be_1$ and $\be_2$, respectively) in its reference configuration, the position vector of the displaced points is 
\begin{equation}
    \br(x) = \Big(x + u(x)\Big)\, \be_1 + v(x)\, \be_2,
\end{equation}
where $u(x)$ and $v(x)$ are the axial and transverse displacement components. 
The tangent to the deformed axis is given by the derivative of $\br$ with respect to $x$,
\begin{equation}
    \br'(x) = \Big(1 + u'(x)\Big)\, \be_1 + v'(x) \be_2,
\end{equation}
so that the stretch of the axis is $\lambda(x) = |\br'(x)|$. 

\begin{figure}[hbt!]
    \centering
    \begin{subfigure}[]{0.80\textwidth}
        \centering
        \includegraphics[width=\textwidth, keepaspectratio]{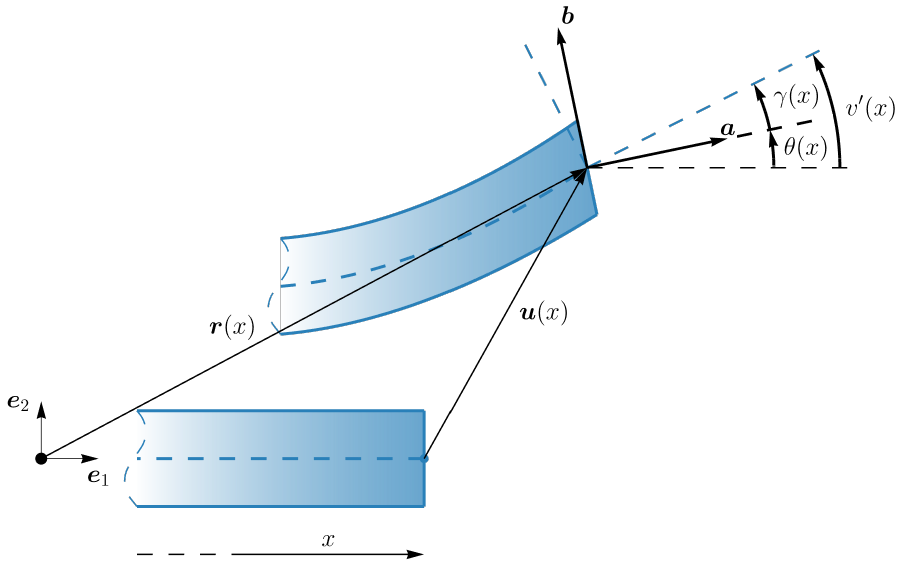}
    \end{subfigure}
    \caption{
    Deformed configuration of the planar Timoshenko beam at coordinate $x$. The material point $\bx(x) = x\,\be_1$ is mapped to its deformed position $\br(x)$. Vectors $\ba$ and $\bb$ are, respectively, orthogonal and tangent to the cross-section at $\br(x)$, as defined in Eq.~\eqref{eq:normal_parallel_cross_section}. The figure also illustrates the slope of the beam axis, $v'(x)$, the rotation of the cross-section, $\theta(x)$, and the shear deformation, $\gamma(x) = v'(x)-\theta(x)$.  
    }
\label{fig:timoshenko_beam}
\end{figure}

The cross-section of the beam rotates at an angle $\gamma$ with respect to the tangent $\br'$, while the normal is inclined at the angle $\theta$, as sketched in Fig.~\ref{fig:timoshenko_beam}. Introducing the unit vectors $\ba$ and $\bb$, respectively normal and parallel to the cross section, 
\begin{equation}
\label{eq:normal_parallel_cross_section}
    \ba = \cos \theta\, \be_1 + \sin \theta\, \be_2, \quad
    \bb = -\sin \theta\,\be_1 + \cos \theta\,\be_2, 
\end{equation}
the tangent $\br'$ can be understood to condensate the effects of the axial $\varepsilon = \lambda - 1$ and shear $\gamma$ strains through the equation
\begin{equation}
\label{cine}
    \be_1 + \bu' = (1+\varepsilon) (\cos\gamma\,\ba + \sin\gamma\,\bb), 
\end{equation}
which upon linearization in $\varepsilon$ and $\gamma$ becomes
\begin{equation}
\label{cine_lin}
    \be_1 + \bu' = (1+\varepsilon)\, \ba + \gamma\, \bb, 
\end{equation}
where $\bu(x) = u(x) \be_1 + v(x) \be_2$ is the vector collecting the axial and transverse displacements.

Under the assumption of small deflection and small rotations, Eq.~\eqref{cine_lin} leads to the linearized equations governing the kinematics of a planar Timoshenko beam
\begin{equation}
\label{cinema}
    \varepsilon(x) = u^{\prime}(x)\,, \quad
    \gamma(x) = v^{\prime}(x) - \theta(x)\,, \quad
    \chi(x) = \theta^{\prime}(x)\,,    
\end{equation}
where the last equation defines the curvature $\chi$ of the beam axis associated with the rotation of the cross-section $\theta$. The linearized equilibrium equations in the absence of distributed forces are
\begin{equation}
    N' = 0, \quad V' = 0, \quad M' = -V, 
\end{equation}
where $N$, $V$, and $M$ are the normal and shear forces and the bending moment. In addition, linear constitutive equations are assumed
\begin{equation}
\label{costi}
    N = E_s A\, \varepsilon, \quad V = \kappa\, G_s A\, \gamma, \quad M = E_s J\, \chi,
\end{equation}
where $E_s$ and $G_s$ are the elastic and shear moduli, $A$ and $J$ the cross-section area and second moment of area, while $\kappa$ is the shear correction factor, introduced by Timoshenko~\cite{timoshenko_shear_1921,timoshenko_shear_1922} to account for the non-uniform distribution of the shear stress across the cross-section. For a rectangular cross-section $\kappa = 5/6$; however, this value is underestimated, and more accurate relationships have been provided~\cite{cowper_shear_coefficient_1966,kaneko_shear_1975,hutchinson_shear_2001,stephen_shear_1980}. In particular, Cowper~\cite{cowper_shear_coefficient_1966} obtained an analytical relation for the shear correction factor as a function of the Poisson's ratio $\nu_s$ of the material, $\kappa(\nu_s)$, by integrating the equations of three-dimensional elasticity, which is suitable for any uniform beam.

Assuming, for simplicity, that all the elastic coefficients are constant and combining equations (\ref{cinema})--(\ref{costi}), the equilibrium equations in terms of the kinematic variables become
\begin{equation}
\label{eq:system_odes}
    u'' = 0, \quad v'' = \theta', \quad  
    \theta''= \frac{\kappa\, G_s A}{E_s J} (\theta-v'), 
\end{equation}
and the elastic strain energy of a Timoshenko beam of length $l$ is defined as
\begin{equation}
    \label{eq:potential_energy}
    \mE = \frac{1}{2}\int_{0}^{l} \Big( E_s A\, u^{\prime\,2} + \kappa\, G_s A\, (v^{\prime} - \theta )^2 + E_s J \, \theta^{\prime\,2} \Big)\,dx\,.
\end{equation}

A comparison between Eqs.~\eqref{eq:potential_energy} and \eqref{discretazzo} reveals that a microstructured rod behaves as a Timoshenko beam, but with $E_s A$, $\kappa\, G_s A$, and $E_s J$ replaced by $k_a a$, $k_t a$, and $k_r a$, respectively. It is important to note that these stiffness components can vary independently due to the microstructural assumptions.

The solution to the system~\eqref{eq:system_odes} can be sought by assuming exponential functions for axial and transverse displacements
\begin{equation}
    \label{eq:general_solution}
    u(x) = \sum_{j=1}^2 C_j \, e^{i \alpha_j x}\,,\quad v(x) = \sum_{k=1}^4 D_k \, e^{i \beta_k x}\,,
\end{equation}
where $i = \sqrt{-1}$ is the imaginary unit,  $\{C_1,C_2,D_1,D_2,D_3,D_4\}$ are six arbitrary constants, and $\{\alpha_1, \alpha_2, \beta_1, \beta_2, \beta_3, \beta_4\}$ are the six characteristic roots. In the present case, all characteristic roots vanish. As a result, the axial and transverse displacements reduce to a linear and a cubic polynomial, respectively:
\begin{equation}
    u(x) = C_1 + C_2\,x, \quad  v(x) = D_1 + D_2\, x + D_3\, x^2 + D_4\, x^3\,,
\end{equation}
where $x \in [0,l]$. In addition, the rotation $\theta$ can be obtained by imposing $\theta'' = v'''$ and solving the equation~$\eqref{eq:system_odes}_3$ for $\theta$, leading to the general solution for the cross-section rotation 
\begin{equation}
    \theta(x) = D_2 + 2 D_3x + 3 D_4\left( x^2 + 2\frac{E_s J}{\kappa\, G_s A} \right).
\end{equation}

The six above constants can be determined by imposing the boundary conditions at the ends of the beam
\begin{equation}
    u(0) = u_0\,,\quad v(0) = v_0\,,\quad \theta(0) = \theta_0\,,\quad u(l) = u_l\,,\quad v(l) = v_l\,,\quad \theta(l) = \theta_l\,,
\end{equation}
which will be collected in the vector $\bq = \trans{[u_0,v_0,\theta_0,u_l,v_l,\theta_l]}$. Consequently, the displacement field $\bu = \trans{[u(x),v(x),\theta(x)]}$ can be written as a function of the vector of nodal degrees of freedom $\bq$ as
\begin{equation}
\label{eq:uNq}
    \bu (x) = \bN(x)\,\bq\,,
\end{equation}
where $\bN (x) = \be_1 \otimes \bN_u (x) + \be_2 \otimes \bN_v (x) + \be_3 \otimes \bN_\theta (x)$ is a $3 \times 6$ matrix determining the exact shape functions of a Timoshenko beam, while $\be_3 = \be_1 \times \be_2$ is the unit vector denoting the out-of-plane axis. 

The three rows of the matrix $\bN$ define the shape functions for the axial and transverse displacement and the rotation of the cross-section, respectively, 
\begin{equation}
\begin{aligned}
    \bN_u (x) &= 
    \begin{bmatrix}
        1 - \dfrac{x}{l} & 0 & 0 & \dfrac{x}{l} & 0 & 0
    \end{bmatrix}, \\[3mm]
    \bN_v (x) &= \dfrac{12x \left(1-\dfrac{x}{l}\right)}{1 + 12\alpha}
    \begin{bmatrix}
        0 & \dfrac{1+12\alpha}{12x} + \dfrac{l-2x}{12 l^2}  &\dfrac{l(1+6\alpha)-x}{12 l} & 0 & \dfrac{\alpha + \dfrac{x}{4 l} - \dfrac{x^2}{6 l^2}}{l-x} & - \dfrac{6\alpha\,l+x}{12l}
    \end{bmatrix}, \\[3mm]
    \bN_\theta (x) &= \dfrac{6\left(1-\dfrac{x}{l}\right)}{1 + 12\alpha}
    \begin{bmatrix}
        0 & - \dfrac{x}{l^2} & \dfrac{1}{6} + 2 \alpha - \dfrac{x}{2l} & 0 & \dfrac{x}{l^2} & \dfrac{x \left(12 \alpha + 3 \dfrac{x}{l} - 2\right)}{6(l-x)}
    \end{bmatrix},
   \end{aligned}
\end{equation}
where the following dimensionless parameters are introduced 
\begin{equation}
\label{costantazze1}
    \alpha = \frac{E_s J}{\kappa\,  G_s A \, l^2} = \frac{2(1+\nu_s)}{\kappa\,\Lambda^2},
    \quad
    \Lambda = l\sqrt{\frac{A}{J}}, 
\end{equation}
in which $\Lambda$ is the slenderness of the beam.

The elastic energy, Eq.~\eqref{eq:potential_energy}, can be expressed in a matrix form by employing Eq.~\eqref{eq:uNq} as
\begin{equation}
\label{eq:potential_energy_q}
    \mE = \frac{1}{2}\, \trans{\bq}\, \bK_{b}\, \bq,
\end{equation}
where the stiffness matrix $\bK_{b}$ for a Timoshenko beam of length $l$ can be expressed as  
\begin{equation}
    \bK_{b} = \int_0^l \trans{\bB(x)}\, \bE\, \bB(x)\, dx,     
\end{equation}
in which $\bB(x)$ denotes the strain-displacement matrix, while $\bE$ denotes the matrix containing the axial, shear, and bending stiffnesses
\begin{equation}
    \bB(x) = 
    \begin{bmatrix}
        \frac{d}{dx} &0            &0            \\
        0            &\frac{d}{dx} &-1           \\
        0            &0            &\frac{d}{dx}
    \end{bmatrix}\, \bN (x),
    \quad
    \bE = 
    \begin{bmatrix}
        E_s A & 0             & 0     \\
        0     & \kappa\, G_s A  & 0     \\
        0     & 0             & E_s J
    \end{bmatrix}.
\end{equation}
The stiffness matrix can explicitly be written as 
\begin{equation}
\label{sega2}
    \bK_{b} = \dfrac{\alpha\,\kappa G_s A}{l (1 + 12\alpha)} 
    \begin{bmatrix}
        \dfrac{1 + 12 \alpha}{\alpha}\,\phi^2 & 0 & 0 & -\dfrac{1 + 12 \alpha}{\alpha}\,\phi^2 & 0 & 0 \\[1.5ex]
        \cdot & 12 & 6 l & 0 & - 12 & 6 l \\[1.5ex]
        \cdot & \cdot & 4 l^2 (1 + 3\alpha) & 0 & -6 l & 2 l^2 (1 - 6\alpha) \\[1.5ex]
        \cdot & \cdot & \cdot &  \dfrac{1 + 12\alpha}{\alpha}\,\phi^2 & 0 & 0 \\[1.5ex]
        \cdot & \cdot & \cdot & \cdot & 12 & -6 l \\[1.5ex]
        \cdot & \cdot & \cdot & \cdot & \cdot & 4 l^2 (1 + 3\alpha)
    \end{bmatrix},
\end{equation}
where $\phi$ is defined as
\begin{equation}
\label{costantazze2}
    \phi = \sqrt{\frac{E_s A}{\kappa\, G_s A}} = \sqrt{\frac{2(1+\nu_s)}{\kappa}}, 
\end{equation}
and denotes the {\it shear coefficient}, a dimensionless parameter that quantifies the ratio between the axial and shear stiffnesses of the Timoshenko beam. For the homogenized version of the discrete chain shown in Fig.~\ref{trave_a_taglio}, the slenderness and shear coefficient can be equivalently expressed in terms of the axial, shear, and rotational spring stiffnesses as follows
\begin{equation}
\label{eq:lambda-phi_chain}
    \Lambda = l\, \sqrt{\frac{k_a}{k_r}}\,, \quad \phi = \sqrt{\frac{k_a}{k_t}}\,,
\end{equation}
where $l$ is the length of the equivalent beam.

\section{Homogenization of grids of Timoshenko beams}
\label{sec:homogenization}
\begin{figure}[hbt!]
    \centering
    \begin{subfigure}[]{0.48\textwidth}
        \centering
        \includegraphics[width=\textwidth, keepaspectratio]{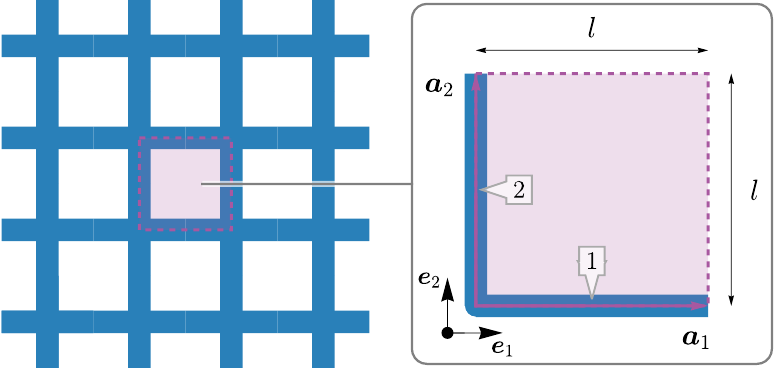}
    \end{subfigure}\hfill
    \begin{subfigure}[]{0.48\textwidth}
        \centering
        \includegraphics[width=\textwidth, keepaspectratio]{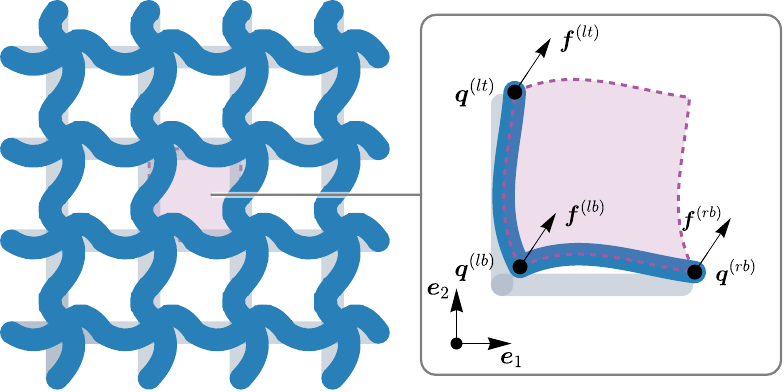}   
    \end{subfigure}
    \caption{
    Undeformed (left) and deformed (right) configurations for a square periodic lattice and its unit cell, made up of Timoshenko beams, represented thick to emphasize the effects of shear deformation. In the right panel, the gray beams represent the undeformed configuration for reference.
    }
\label{fig:generic_undeformed_geometry}
\end{figure}

The static homogenization procedure matches the strain energies to derive an elastic continuum equivalent to a periodic grid of Timoshenko beams, following an approach similar to that applied to Euler-Bernoulli rods \cite{bordiga_tensile_2022}. To this end, a unit cell composed of Timoshenko beams is introduced, as shown in Fig.~\ref{fig:generic_undeformed_geometry}. This cell tessellates a two-dimensional periodic lattice along the directions defined by the direct basis vectors $\ba_1$ and $\ba_2$. Although illustrated as a square for simplicity, the unit cell can have an arbitrary shape. The unit cell shown in Fig.~\ref{fig:generic_undeformed_geometry} for the square lattice is chosen to minimize the number of degrees of freedom in the system.

\paragraph{Elastic strain-energy density of the unit cell.}
On introducing the vectors $\bq$ and $\bef$, collecting the nodal displacements and the dual nodal forces, respectively, the equilibrium of the unit cell can be expressed as 
\begin{equation}
\label{eq:unit_cell_equilibrium_01}
    \bK\, \bq = \bef,
\end{equation}
where $\bK$ is the symmetric stiffness matrix of the unit cell, obtained by assembling the stiffness matrices $\bK_{b}$, Eq.~\eqref{sega2}. Periodicity of the lattice requires for each pair of nodes $\{p,q\}$ such that $\bx_q - \bx_p = n_1 \ba_1 + n_2 \ba_2$, $n_1, n_2 \in \{0,1\}$ that the components of displacement $\tilde \bq^{(j)} = \trans{\begin{bmatrix} \tilde u^{(j)} & \tilde v^{(j)} & \tilde \theta^{(j)} \end{bmatrix}}$ satisfy 
\begin{equation}
    \tilde u^{(q)} = \tilde u^{(p)}, \quad 
    \tilde v^{(q)} = \tilde v^{(p)}, \quad
    \tilde{\theta}^{(q)} = \tilde{\theta}^{(p)}.
\end{equation}
Multiple-scale asymptotic expansion of displacement field for periodic composites \cite{bensoussan_asymptotic, cioranescu} suggests that, in addition to a periodic field (generating a vanishing macroscopic strain), an affine deformation $\bm \epsilon\,\bx^{(j)}$ is considered
\cite{hutchinson_structural_2006}, so that the total displacement field can be written as
\begin{equation}
    \bq^{(j)} =
    \trans{\begin{bmatrix}
        \tilde u^{(j)} + (\bm \epsilon \, \bx^{(j)}) \cdot \be_1 & \tilde v^{(j)} + (\bm \epsilon \, \bx^{(j)}) \cdot \be_2 & \tilde \theta^{(j)}
    \end{bmatrix}},
\end{equation}
where $\bm\epsilon$ is a given uniform strain tensor (representing a homogeneous strain for the unit cell) and $\bx^{(j)}$ is the vector collecting the coordinates of the $j$-th node.

With reference to the unit cell reported in Fig.~\ref{fig:generic_undeformed_geometry}, vectors $\bq$ and $\bef$ can be partitioned as
\begin{equation}
    \bq = 
    \trans{
    \begin{bmatrix}
        \bq^{(lb)} & \bq^{(rb)} & \bq^{(lt)}
    \end{bmatrix}
    },
    \qquad
    \bef = 
    \trans{
    \begin{bmatrix}
        \bef^{(lb)} & \bef^{(rb)} & \bef^{(lt)}
    \end{bmatrix}}.
\end{equation}

Periodicity implies that the vectors $\tilde\bq^{(rb)}$, $\tilde\bq^{(lt)}$ can be expressed through the \lq left-bottom' term $\tilde\bq^{(lb)}$ as
\begin{equation}
    \tilde \bq = 
    \begin{bmatrix}
        \tilde \bq^{(lb)} \\ \tilde \bq^{(rb)} \\ \tilde \bq^{(lt)}
    \end{bmatrix} = 
    \begin{bmatrix}
        \bI \\ \bI \\ \bI
    \end{bmatrix}\,\tilde \bq^{(lb)} = \bZ_0 \, \tilde \bq^*.
\end{equation}
Consequently, the generalized displacement vector $\bq$ can be written as the sum of  the periodic component, $\tilde \bq$, and the homogeneous component, $\hat \bq (\bm \epsilon)$, as follows
\begin{equation}
\label{eq:generalized_displacement_vector}
    \bq (\tilde \bq^*, \bm \epsilon) =  \bZ_0\,\tilde \bq^* + \hat \bq (\bm \epsilon)\,,
\end{equation}
where the homogeneous displacement vector of the $j$-th node is written as
\begin{equation}
    \hat{\bq}^{(j)} (\bm \epsilon)  = 
    \begin{bmatrix}
        (\bm \epsilon \, \bx^{(j)}) \cdot \be_1 \\
        (\bm \epsilon \, \bx^{(j)}) \cdot \be_2 \\
        0
    \end{bmatrix}.
\end{equation}
Finally, Eq.~\eqref{eq:generalized_displacement_vector} can be substituted into the equilibrium equation~\eqref{eq:unit_cell_equilibrium_01} for the unit cell, so that a pre-multiplication by $\trans{\bZ}_0$ leads to the following expression 
\begin{equation}
\label{eq:unit_cell_equilibrium_02}
    \trans{\bZ}_0 \, \bK \, \bZ_0\,\tilde\bq^* + \trans{\bZ}_0 \, \bK \, \hat \bq (\bm \epsilon) = \trans{\bZ}_0 \, \bef\,.
\end{equation}
The right-hand side term can be represented as the sum of three nodal force vectors
\begin{equation}
    \trans{\bZ}_0 \, \bef = \bef^{(lb)} + \bef^{(rb)} + \bef^{(lt)} ,
\end{equation}
which vanishes in the absence of external loads, as the internal forces shown in Fig.~\ref{fig:generic_undeformed_geometry} satisfy the equilibrium condition, $\bef^{(lb)} = - \bef^{(rb)} - \bef^{(lt)}$. Consequently, Eq.~\eqref{eq:unit_cell_equilibrium_02} reduces to
\begin{equation}
\label{eq:unit_cell_equilibrium_03}
   \trans{\bZ}_0\, \bK\, \bZ_0\, \tilde\bq^* = -\trans{\bZ}_0 \, \bK \, \hat \bq (\bm \epsilon).
\end{equation}
The stiffness matrix $\bK$ is rank-deficient because the unit cell is unconstrained, allowing three independent rigid-body motions. Applying the periodicity conditions, represented by $\bZ_0$, suppresses the rigid-body rotation while still permitting two rigid-body translations. As a result, for any vector $\bt$ representing a pure rigid-body translation, the following relation holds
\begin{equation}
\label{eq:rigid}
    \bK\, \bZ_0\, \bt = \bm 0,
\end{equation}
which shows that $\dim \Bigl(\ker( \trans{\bZ_0} \, \bK \, \bZ_0)\Bigr) \geq 2$. The nullspace of $\trans{\bZ_0} \, \bK \, \bZ_0$ therefore contains \textit{at least} two zero-energy modes associated with rigid-body translations. Any additional zero-energy mode in $\ker{(\trans{\bZ_0} \, \bK \, \bZ_0)}$ is referred to as an internal mechanism, or \lq floppy mode'. Floppy modes are excluded in the present treatment.

The \textit{Fredholm alternative theorem} states that the system~\eqref{eq:unit_cell_equilibrium_03} admits a solution if and only if the right-hand side, $-\trans{\bZ}_0 \, \bK \, \hat \bq (\bm \epsilon)$, lies in the orthogonal complement of the nullspace of the coefficient matrix $\trans{\bZ}_0\, \bK\, \bZ_0$, that is
\begin{equation}
    -\trans{\bZ}_0 \, \bK \, \hat \bq (\bm \epsilon) \in \ker{(\trans{\bZ_0} \, \bK \, \bZ_0)}^\perp.
\end{equation}
This condition is always satisfied, since any vector $\bt$ in $\ker{(\trans{\bZ_0} \, \bK \, \bZ_0)}$ represents a rigid-body translation. As a result, 
\begin{equation}
    \bt \cdot \trans{\bZ}_0\, \bK\, \hat\bq(\bm \epsilon) = \bK\, \bZ_0\, \bt \cdot \hat\bq(\bm \epsilon) = 0,
\end{equation}
as implied by Eq.~\eqref{eq:rigid}.
The solution of system~\eqref{eq:unit_cell_equilibrium_03} can therefore obtained as a linear function of $\bm \epsilon$
\begin{equation}
    \tilde \bq^* = \tilde \bq^* (\bm \epsilon).
\end{equation}
Consequently, the elastic strain-energy density of the unit cell takes the form of a quadratic function in $\bm \epsilon$ expressed as
\begin{equation}
\label{eq:elastic_lattice_strain_energy}
    \mE (\bm\epsilon) = \frac{1}{2|\mC|} \, \bq (\tilde \bq^*(\bm \epsilon), \bm \epsilon) \cdot \bK \, \bq (\tilde \bq^*(\bm \epsilon), \bm \epsilon),
\end{equation}
where $|\mC|$ denotes the area of the unit cell.

\paragraph{Matching with the elastic strain-energy density of an elastic continuum.}
\label{sec:homogenization_technique}
The homogenization process establishes a connection between the microscopic behavior of the lattice and the macroscopic response of an equivalent Cauchy elastic continuum by imposing the equivalence of their elastic strain-energy densities. The strain-energy density of the unit cell corresponds to a deformation imposed through nodal displacements, while the strain-energy density of the equivalent continuum is given by
\begin{equation}
\label{eq:elastic_continuum_strain-energy}
    \mW(\bm \epsilon) = \frac{1}{2}\, \bm \epsilon \cdot \fE[\bm \epsilon]\,,
\end{equation}
with the associated stress-strain relation $\bm \sigma = \fE [ \bm \epsilon ]$. Enforcing the equivalence between the continuum and lattice elastic strain energies, Eqs.~\eqref{eq:elastic_lattice_strain_energy} and \eqref{eq:elastic_continuum_strain-energy}, yields
\begin{equation}
\label{eq:equivalence_strain_energies}
    \bm \epsilon \cdot \fE[\bm \epsilon] = \frac{1}{|\mC|} \, \bq (\tilde \bq^*(\bm \epsilon), \bm \epsilon) \cdot \bK \, \bq (\tilde \bq^*(\bm \epsilon), \bm \epsilon).
\end{equation}
The elasticity tensor, which characterizes the effective continuum material, follows from relation~\eqref{eq:equivalence_strain_energies} as the second derivative of the elastic strain-energy density with respect to the macroscopic strain $\bm \epsilon$, as 
\begin{equation}
\label{eq:elasticity_tensor}
    \fE = \frac{1}{2 |\mC|} \, \nderiv{\Bigl(\bq (\tilde \bq^*(\bm \epsilon), \bm \epsilon) \cdot \bK \, \bq (\tilde \bq^*(\bm \epsilon), \bm \epsilon)\Bigr)}{\bm \epsilon}{2}\,,
\end{equation}
which exhibits all the required symmetries \cite{bigoni_nonlinear_2012, gurtin1982introduction}.

\section{Influence of the beam shear stiffness on the effective elastic properties}
\label{sec:influence_shear}
This section investigates how the shear stiffness of the beams forming the grid (assumed infinite in the Euler-Bernoulli model but finite in the Timoshenko model) affects the behavior of the equivalent elastic material.

For a two-dimensional Cauchy elastic material, the constitutive law can be expressed in Voigt notation and with respect to an orthonormal basis $\be_i$ ($i=1,2$) as
\begin{equation}
\label{eq:elastic_constitutive_law}
    \begin{bmatrix}
        \sigma_{11} \\
        \sigma_{22} \\
        \sigma_{12}
    \end{bmatrix}
    =
    \begin{bmatrix}
        \fE_{1111}  & \fE_{1122}      & \fE_{1112} \\
        \cdot       & \fE_{2222}      & \fE_{2212} \\
        \cdot       & \cdot           & \fE_{1212} \\
    \end{bmatrix}
    \begin{bmatrix}
        \epsilon_{11} \\
        \epsilon_{22} \\
        2\epsilon_{12}
    \end{bmatrix}.
\end{equation}
Since the elasticity tensor $\fE$ is positive definite, Eq.~\eqref{eq:elastic_constitutive_law} can be inverted as $\bm \epsilon = \fE^{-1}[\bm \sigma]$, where $\fE^{-1}$ is the compliance tensor. In Voigt notation, under the general case of material anisotropy and assuming a state of {\it plane stress}, the inverse relation takes the form
\begin{equation}
\label{eq:inverse_constitutive_law}
    \begin{bmatrix}
        \epsilon_{11} \\
        \epsilon_{22} \\
        2\epsilon_{12}
    \end{bmatrix}
    = 
    \begin{bmatrix}
        \dfrac{1}{E_1} & -\dfrac{\nu_{21}}{E_2} & \dfrac{\eta_{1,12}}{G_{12}} \\[1.5ex]
        -\dfrac{\nu_{12}}{E_1} & \dfrac{1}{E_2} & \dfrac{\eta_{2,12}}{G_{12}} \\[1.5ex]
         \dfrac{\eta_{12,1}}{E_{1}} & \dfrac{\eta_{12,2}}{E_{2}} & \dfrac{1}{G_{12}}
    \end{bmatrix}
    \begin{bmatrix}
        \sigma_{11} \\
        \sigma_{22} \\
        \sigma_{12}
    \end{bmatrix},
\end{equation}
where the compliance tensor includes the Young's moduli $E_1$ and $E_2$, the shear modulus $G_{12}$, the Poisson's ratios, $\nu_{12}$ and $\nu_{21}$, and the mutual influence coefficients of the first and second kind, defined as
\begin{equation}
    \eta_{i,ij} = \eta_{i,ji} =  \frac{\epsilon_{ii}}{2\epsilon_{ij}}  =  \frac{\epsilon_{ii}}{2\epsilon_{ji}}\,,
    \qquad \eta_{ij,i}=\eta_{ji,i} = \frac{2\epsilon_{ij}}{\epsilon_{ii}}= \frac{2\epsilon_{ji}}{\epsilon_{ii}}\,, \qquad i,j=1,2 \quad
    \text{ not summed}.
\end{equation}
The coefficients $\eta_{i,ij}$ represent the ratio of normal to tangential strain components under pure shear stress, while $\eta_{ij,i}$ correspond to the inverse of the strain ratio under a pure normal stress applied in the directions 1 and 2~\cite{lekhnitskii_anistropic_1964}, respectively.

The following analysis adopts dimensionless geometrical and mechanical parameters to describe each unit cell. Specifically, the $N_b$ beams forming the unit cell are characterized by their slenderness $\Lambda$ and shear coefficient $\phi$. For a three-dimensional, linear elastic, isotropic material, the Poisson's ratio $\nu_s$ is restricted to the interval $(-1, 1/2)$, which implies that $\phi \in (0, \sqrt{3/\kappa})$. For example, assuming a rectangular cross-section, where $\kappa = 5/6$, and a typical Poisson's ratio of $\nu_s = 0.3$, the corresponding shear coefficient is approximately $\phi \simeq 1.76$. 

For the sake of simplicity, all beams are assumed to share the same shear coefficient $\phi$, as defined in Eq.~$\eqref{costantazze2}$. In the context of Euler-Bernoulli beam theory, the shear stiffness $\kappa\, G_s A$ is taken as infinite, which corresponds to the limiting case $\phi \to 0$. This condition also occurs when the Poisson's ratio of the material tends to the lower bound, $\nu_s \to -1$, leading to $G_s \to +\infty$. In this limit, the Euler-Bernoulli and Timoshenko beam models coincide.

In the following, the relationship between the elastic properties of the effective continuum and the slenderness of the beams is investigated through the determination of analytical expressions for the elasticity tensor characterizing the equivalent continuum, as functions of the shear coefficient $\phi$ and slenderness $\Lambda$.
It is important to note that the influence of shear deformability is assessed under the assumption of a homogeneous elastic beam, where axial, shear, and bending stiffnesses are intrinsically linked. However, by employing more advanced models of microstructured rods \cite{Kocsis, paradiso}, it is possible to decouple shear stiffness from bending stiffness, allowing for independent control of these properties, as explained in Section~\ref{sec:decoupled_plane_stress}.

\subsection{Square lattice and the stubby-beam limit for a cubic equivalent material}
\label{sec:square_lattice}
The square grid shown in Fig.~\ref{fig:generic_undeformed_geometry} is generated by tessellating a square unit cell composed of two {\it identical} elastic Timoshenko beams of length $l$. The dimensionless parameters characterizing these beams are
\begin{equation}
\label{melone}
    \chi = \frac{A_2}{A_1}=1\,, \quad 
    \Lambda_1 = \Lambda_2 = \Lambda\,, \quad 
    \phi_1 = \phi_2 = \phi\,,
\end{equation}
where $A_i$ denotes the cross-section area of the $i$-th beam, with subscripts 1 and 2 corresponding to horizontal and vertical beams, respectively. Under these assumptions, the elasticity tensor in Voigt notation becomes diagonal, with $\fE_{1111} = \fE_{2222}$. 

\begin{figure}[hbt!]
    \centering
    \begin{subfigure}[]{.96\textwidth}
        \centering
        \includegraphics[width=\textwidth]{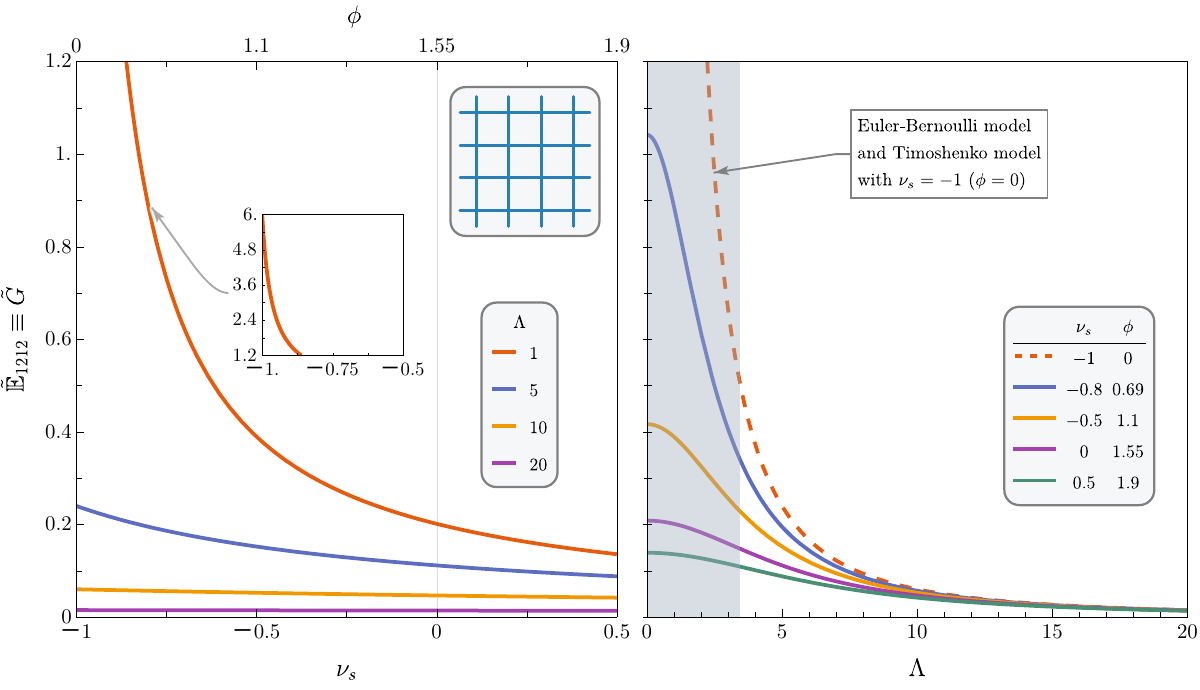}
    \end{subfigure}
    \caption{
    Effective shear modulus $\tilde{G}=\tilde{\fE}_{1212}$ for a square grid of Timoshenko beams with rectangular cross-section, plotted as a function of the Poisson's ratio $\nu_s$ of the beam material (left) and as a function of the beam slenderness $\Lambda$ (right). In the limit of stubby beams ($\Lambda \to 0$), the Timoshenko model predicts a finite shear stiffness for $\nu_s \neq -1$, whereas the Euler-Bernoulli model (red dashed line) leads to an infinite value. The shear stiffness increases as $\nu_s$ becomes more negative, reaching a maximum in the limit $\nu_s \to -1$. 
    }
    \label{fig:square_lattice_G12-nus_lambda}
\end{figure}

By introducing an out-of-plane thickness $h$ for the equivalent solid, taken as unity for plane strain or as the thickness of the equivalent plate for plane stress, a dimensionless elasticity tensor can be defined as  
\begin{equation}
\label{peretta}
    \tilde{\fE} = \frac{l\, h}{E_s A}  \, \fE ,
\end{equation}
so that the {\it non-zero} elastic constants characterizing the equivalent material are given by
\begin{equation}
    \tilde\fE_{1111} = \tilde\fE_{2222} = 1, \quad 
    \tilde\fE_{1212} = \frac{6}{\Lambda^2 + 12 \phi^2}.
\end{equation}
In Fig.~\ref{fig:square_lattice_G12-nus_lambda}, the variation of the non-trivial elastic constant, $\tilde{G}=\tilde{\fE}_{1212}$, is examined. The left panel illustrates its dependence on the Poisson's ratio $\nu_s$ of the isotropic material forming the beams, which are assumed to have a rectangular cross-section, so that $\kappa=5/6$ in Eq.~\eqref{costantazze2}. The right panel explores the effect of beam slenderness $\Lambda$ on $\tilde{\fE}_{1212}$. In the right part of the figure, the Timoshenko beam model with $\nu_s=-1$ aligns with the Euler-Bernoulli model, as evidenced by the red dashed line.

It is worth noting that the influence of the constituent material reduces as the beam becomes more slender and becomes negligible for $\Lambda>15$. As the slenderness approaches the theoretical limit $\Lambda \to +\infty$, the grid behavior converges to that of a hinged square lattice, consistently with the definition of $\Lambda$, Eq.~$\eqref{costantazze1}$. In this limit, the equivalent material retains the capacity to withstand uniaxial and biaxial stresses along the principal directions $\be_1$ and $\be_2$, since $\tilde{\fE}_{1111}$ and $\tilde{\fE}_{2222}$ are independent of $\Lambda$. However, the shear stiffness vanishes, $\fE_{1212} \to 0$, indicating a loss of resistance to shear deformation.   

Although the results are plotted over the full range of slenderness $\Lambda$, values smaller than, say, 4 are generally unrealistic for conventional structural elements. However, small values of $\Lambda$ are possible in the case of microstructured beams. It is therefore of interest to examine the asymptotic behavior as $\Lambda \to 0$, where the Timoshenko model predicts a finite shear stiffness, while the Euler-Bernoulli model leads to an infinite value. Thus, the Timoshenko model corrects the tendency of the Euler-Bernoulli formulation to predict infinite shear stiffness at vanishing slenderness. This highlights the limitations of the Euler-Bernoulli theory for low-slenderness beams, whereas the Timoshenko model remains valid and physically meaningful.

\subsection{Lattices with isotropic effective properties}
Only tessellations based on equilateral triangles or regular hexagons (honeycombs) yield isotropy in the equivalent continuum. 
In a two-dimensional continuum theory, there are only four symmetry classes: triclinic (or rhombic), orthotropic (or rectangular), cubic (or square), and isotropic. In 2D, the lattices that tessellate the plane are oblique, rectangular, cubic, and hexagonal. The hexagonal case can be based either on equilateral triangles --forming a Bravais lattice-- or on a honeycomb pattern, which is not a simple lattice. Hexagonal lattices introduce six symmetry axes (rotational and reflective). To preserve these symmetries in the equivalent continuum, the continuum must be isotropic. Except in special cases of constitutive behavior, the other lattices do not possess sufficient symmetry to correspond to an isotropic continuum. It should also be noted that applying higher-order homogenization (not considered here) may lead to a loss of symmetry in the equivalent continuum \cite{bacca}. 

Hexagonal geometries, which are considered in this section, result in an isotropic elastic tensor that satisfies the following constraints
\begin{equation}
\label{eq:isotropy_elasticity_components}
    \fE_{1111} = \fE_{2222}\,,\quad \fE_{1212} = \frac{\fE_{1111} - \fE_{1122}}{2}\,,\quad \fE_{1112}=\fE_{2212} = 0\,.
\end{equation}
Referring to a purely two-dimensional and isotropic continuum, where 
\begin{equation}
    \sigma_{\alpha \beta} = 
    \ell\, \varepsilon_{\gamma \gamma}\, \delta_{\alpha \beta} + 2m\, \varepsilon_{\alpha \beta}, \quad \alpha, \beta, \gamma = 1,2, 
\end{equation}
in which $\ell$ and $m$ play the role of 2D-Lam\'e constants, the elasticity tensor is written as
\begin{equation}
    \fE =  \begin{bmatrix}
        \ell+2m & \ell      & 0 \\
        \ell    & \ell+2m   & 0 \\
        0       & 0         & m
    \end{bmatrix},
\end{equation}
and the 2D Poisson's ratio $n$ and elastic modulus $e$ are defined as 
\begin{equation}
    n = \frac{\ell}{\ell+2m}, \quad e = 2(1+n)m,
\end{equation}
respectively. Positive definiteness of the elastic energy for a 2D continuum requires
\begin{equation}
    \ell+2m>0, \quad m>0, \quad e>0, \quad -1 < n < 1,
\end{equation}
so that for 
\begin{equation}
    \begin{array}{lllll} 
    \text{plane strain } & 
    \displaystyle \ell = \frac{2\lambda \mu}{\lambda-2\mu}, & m = \mu, 
    & \displaystyle e=\frac{E}{1-\nu^2}, & \displaystyle n =\frac{\nu}{1-\nu}, \\[5mm]
    \text{plane stress } & 
    \ell = \lambda, & m = \mu, & e = E, & n = \nu, 
    \end{array}
\end{equation}
where $\lambda$ and $\mu$ are the Lam\'e constants and $E$ and $\nu$ the elastic modulus and the Poisson's ratio of the 3D theory of elasticity. 

\begin{figure}[hbt!]
    \centering
    \begin{subfigure}[]{.48\textwidth}
        \centering
        \includegraphics[width=\textwidth]{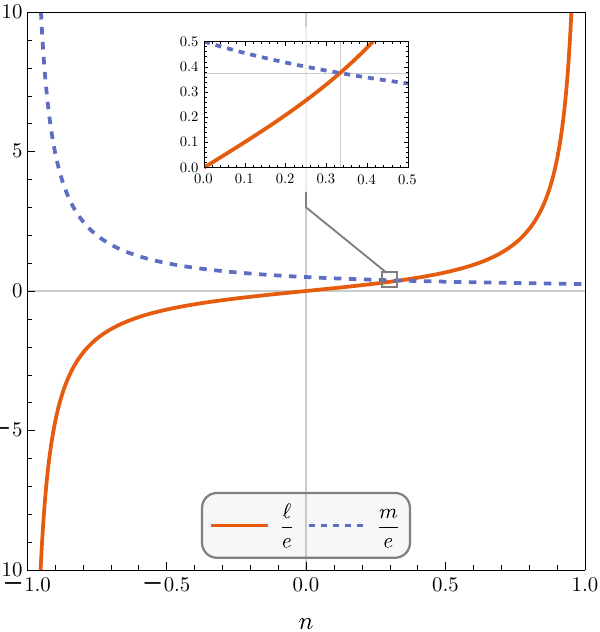}
    \end{subfigure}
    \caption{2D-Lamé constants $\ell$ and $m$ (made dimensionless through division by the elastic modulus $e$) as functions of the Poisson's ratio $n \in (-1,1)$, for a two-dimensional isotropic material.}
    \label{fig:2d_lame_constants-n}
\end{figure}

As the 2D-Poisson's ratio $n$ approaches the lower (the upper) limit $-1$ ($+1$), the shear modulus $m$ tends to infinity (tends to $e/4$), as represented in Fig.~\ref{fig:2d_lame_constants-n}.  

In the following, three different lattices with isotropic effective properties are analyzed. 

\subsubsection{Triangular lattice and the false auxeticity arising from Euler-Bernoulli beam modeling}
\label{sec:triangular_lattice}
\begin{figure}[hbt!]
    \centering
    \begin{subfigure}[]{0.6\textwidth}
        \centering
        \includegraphics[width=\textwidth, keepaspectratio]{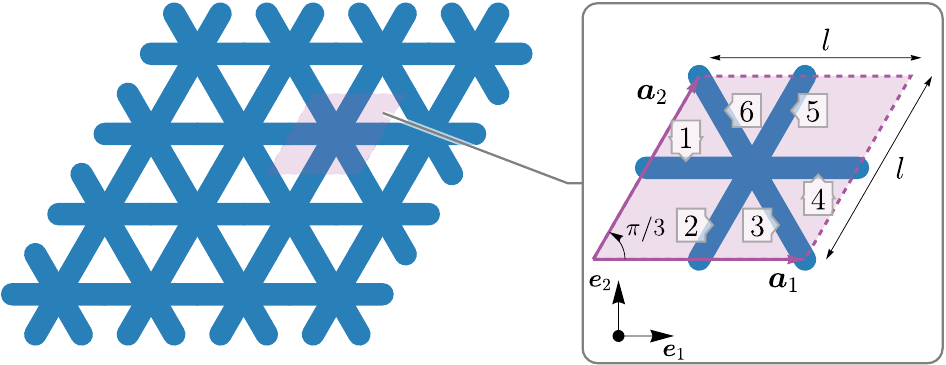}
    \end{subfigure}
    \caption{
    Triangular beam lattice (left) and corresponding unit cell composed of six elastic Timoshenko beams (right).
    }
    \label{fig:triangular_elastic_lattice_geometry}
\end{figure}

A triangular lattice composed of Timoshenko beams is analyzed, with a unit cell consisting of six beams, all of equal length $l/2$, as illustrated in Fig.~\ref{fig:triangular_elastic_lattice_geometry}. 

All beams have the same cross-sectional area $A$, moment of inertia $J$, slenderness $ \Lambda$, and shear coefficient $\phi$. The elastic constants of the continuum material equivalent to the grid are normalized according to equation~\eqref{peretta} and depend on the parameters $\phi$ and $\Lambda$ as follows
\begin{equation}
\label{eq:triangular_lattice_elasticity_tensor}
    \begin{bmatrix}
        \tilde\fE_{1111} & \tilde\fE_{1122} & \tilde\fE_{1112} \\
        \cdot & \tilde\fE_{2222} & \tilde\fE_{2212} \\
        \cdot & \cdot & \tilde\fE_{1212}
    \end{bmatrix} = 
    \frac{3\sqrt{3}(\Lambda^2+12\phi^2+4)}{4(\Lambda^2+12\phi^2)}
    \begin{bmatrix}
        1 & \frac{\Lambda^2+12(\phi^2-1)}{3(\Lambda^2+12\phi^2+4)} & 0 \\
        \cdot & 1 & 0 \\ 
        \cdot & \cdot & \frac{\Lambda^2+12(\phi^2+1)}{3(\Lambda^2+12\phi^2+4)}
    \end{bmatrix}.
\end{equation}
The inverse of the elasticity tensor for the equivalent material, Eq.~\eqref{eq:triangular_lattice_elasticity_tensor}, yields the compliance tensor $\fE^{-1}$, which, for a state of plane stress, can be expressed in terms of two independent constants as 
\begin{equation}
\label{eq:triangular_lattice_compliance_tensor}
    \tilde{\fE}^{-1} = 
    \begin{bmatrix}
        \dfrac{1}{\tilde E} & -\dfrac{\nu}{\tilde E}  & 0 \\[1.5ex]
        \cdot & \dfrac{1}{\tilde E} & 0 \\[1.5ex]
        \cdot & \cdot & \dfrac{2(1+\nu)}{\tilde E}
    \end{bmatrix},
\end{equation}
where $\tilde E = E\, l\, h/ (E_s A)$ and $\nu$ denote the dimensionless Young's modulus and the Poisson's ratio, respectively, depending on $\phi$ and $\Lambda$ as
\begin{equation}
\label{eq:triangular_lattice_engineering_constants}
    \tilde E = \frac{2(\Lambda^2 + 12 (\phi^2 + 1))}{\sqrt{3}(\Lambda^2 + 12 \phi^2 + 4)}, \quad 
    \nu = \frac{\Lambda^2 + 12 (\phi^2 - 1)}{3(\Lambda^2 + 12 \phi^2 + 4)}.
\end{equation}

\begin{figure}[hbt!]
    \centering
    \begin{subfigure}[]{.96\textwidth}
        \centering
        \includegraphics[width=\textwidth]{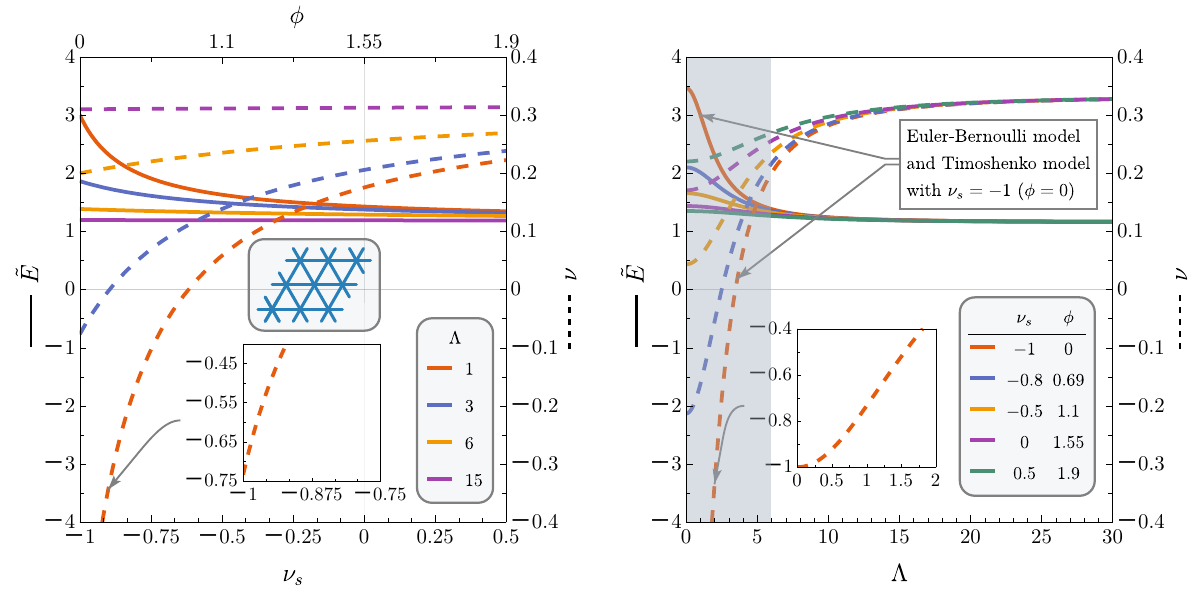}
    \end{subfigure}
    \caption{
    Effective Young's modulus $\tilde E$ (continuous curves) and Poisson's ratio $\nu$ (dashed curves) for the triangular lattice, shown as functions of the Poisson's ratio $\nu_s$ (and $\phi$) of the beam material (left) and of the beams slenderness $\Lambda$ (right). Note that the Euler-Bernoulli model coincides with the Timoshenko beam model when $\phi = 0$; in this case, the effective Poisson's ratio reaches its minimum value, $\nu=-1$, as $\Lambda \to 0$ (right). 
    }
    \label{fig:triangular_elastic_lattice_E_nu}
\end{figure}

In Fig.~\ref{fig:triangular_elastic_lattice_E_nu}, Eqs.~\eqref{eq:triangular_lattice_engineering_constants} are plotted assuming that the beams have a rectangular cross-section ($\kappa=5/6$) and are made of a linear elastic material, see Eq.~\eqref{costantazze2}. In the left panel of the figure, $\tilde E$ (continuous curves) and $\nu$ (dashed curves) are shown as functions of the Poisson's ratio of the beam material, $\nu_s$ (or parameter $\phi$), for four values of slenderness. In the right panel of the figure, $\tilde E$ (continuous curves) and $\nu$ (dashed curves) are plotted as functions of the slenderness $\Lambda$ for five values of $\nu_s$ or $\phi$. 

It is worth noting that for beams made of a material with a negative Poisson's ratio within the range $\nu_s \in (-0.618,0)$, the effective continuum material exhibits a positive Poisson's ratio regardless of the slenderness. Conversely,  when the Poisson's ratio of the beam material lies within $\nu_s \in (-1, -0.618)$, the equivalent continuum material of the triangular lattice exhibits auxetic behavior for sufficiently small slenderness. The most pronounced auxetic effect occurs at the limiting value $\nu_s = -1$ for non-zero slenderness, resulting in an effective Poisson's ratio of $-11/15 \approx -0.733$. 

It can be concluded that, for the triangular lattice under analysis, the absolute value of the effective Poisson's ratio is reduced compared to that of the constituent material, at least for stubby beams. The Poisson's ratio of the constituent material $\nu_s$ can never be exactly attained, except in the limiting case $\nu_s=-1$, which corresponds to the Euler-Bernoulli beam model, when the slenderness approaches its theoretical limit $\Lambda = 0$ (red dashed line in Fig.~\ref{fig:triangular_elastic_lattice_E_nu}, right). However, this limit can be approached only by introducing a microstructured beam. 

On the other hand, the effective Young's modulus, $\tilde E$, increases as $\nu_s$ and $\Lambda$ decrease, reaching its maximum when the Euler-Bernoulli beam model is considered (or when $\nu_s = -1$ in the Timoshenko model) in the limit $\Lambda \to 0$, yielding $\tilde E = 2\sqrt{3} \approx 3.46$. However, this value decreases as $\nu_s$ increases; for example, $\tilde E \approx 1.38$ for $\nu_s = 0.3$ and $\Lambda = 0$).

Note also that the effects of shear deformability vanish as the beams become slender, so that the Timoshenko and Euler-Bernoulli models coincide regardless of $\nu_s$ in the limit $\Lambda \to \infty$, as shown in Fig.~\ref{fig:triangular_elastic_lattice_E_nu} (right).

\begin{figure}[hbt!]
    \centering
    \begin{subfigure}[]{.48\textwidth}
        \centering
        \includegraphics[width=\textwidth]{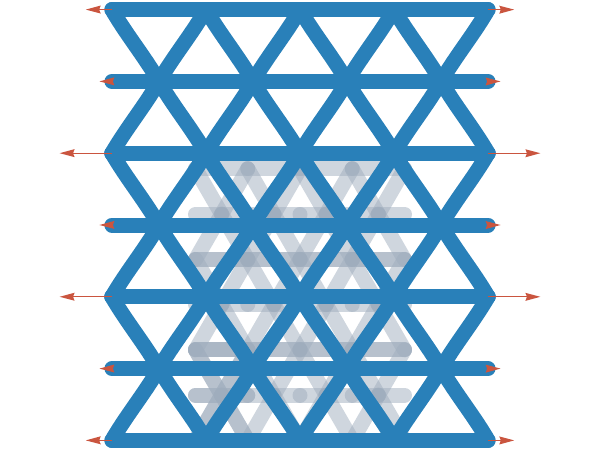}
        \caption{Euler-Bernoulli beam model with $\Lambda=1$, yielding $\nu=-11/15 \approx -0.733$}
    \end{subfigure}
    \begin{subfigure}[]{.48\textwidth}
        \centering
        \includegraphics[width=\textwidth]{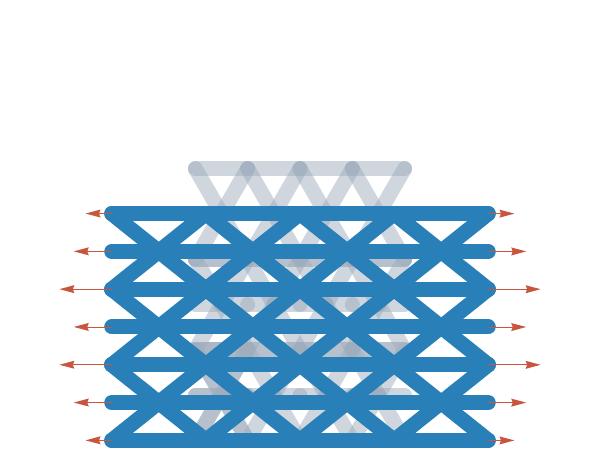}
        \caption{Timoshenko beam model with $\Lambda=1$ and $\nu_s=0.3$, yielding $\nu=0.207$}
    \end{subfigure}
    \caption{
    Deformed triangular lattice under horizontal uniaxial stress with fixed low slenderness $\Lambda=1$, used here to magnify the effect. 
    (a) The Euler-Bernoulli beam model exhibits false auxeticity, with an effective negative Poisson's ratio $\nu = -11/15$ for the equivalent material.
    (b) This effect is corrected by the Timoshenko beam model, which for $\nu_s = 0.3$ yields the correct positive value of the effective Poisson's ratio $\nu = 0.207$. 
    Note that the displacements are shown to scale, while the beam thickness has been reduced for graphical clarity. Red arrows represent nodal forces, with lengths proportional to their magnitude.
    }
    \label{fig:triangular_elastic_deformed}
\end{figure}

A deformed portion of the grid under horizontal uniaxial stress is shown in Fig.~\ref{fig:triangular_elastic_deformed}. On the left, the beams are modeled using the Euler-Bernoulli theory and exhibit false auxeticity, while on the right, the beams (with $\nu_s=0.3$) are modeled using the Timoshenko theory, which provides the correct behavior. 

\subsubsection{Hexagonal elastic lattice with essential bending effects}
\label{sec:hexagonal_lattice}
The hexagon is the polygon with the largest number of sides that can tile the plane without leaving voids and, consequently, minimizes the total perimeter for a given area. This geometric optimization problem, known as the \textit{honeycomb conjecture}, was proven only recently by Hales \cite{hales_honeycomb_2001}. As a result, hexagonal lattices have been extensively studied as models for both two-dimensional \cite{evans_auxetic_1991, evans_design_1991, gibson_cellular_1997, masters_models_1996} and three-dimensional honeycombs \cite{gibson_mechanics_1982}. 

\begin{figure}[hbt!]
    \centering
    \begin{subfigure}[]{0.6\textwidth}
        \centering
        \includegraphics[width=\textwidth, keepaspectratio]{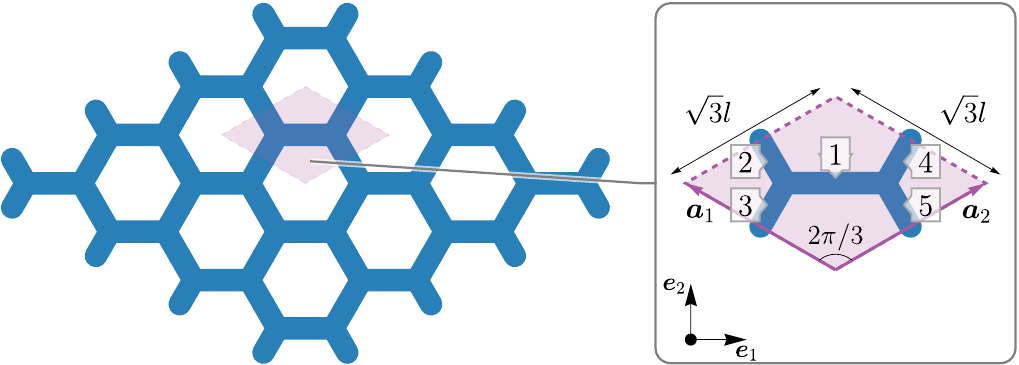}
    \end{subfigure}
    \caption{
    Hexagonal grid (left) generated by tessellating a unit cell composed of five elastic Timoshenko beams (right).
    }
    \label{fig:hexagonal_elastic_lattice_geometry}
\end{figure}

The hexagonal lattice, shown on the left in Fig.~\ref{fig:hexagonal_elastic_lattice_geometry}, is generated by tessellating the unit cell depicted on the right along the axes $\{\ba_1, \ba_2\}$, which form the direct basis. The unit cell consists of two V-shaped elements, each composed of two identical elastic Timoshenko beams of length $l/2$, connected by a horizontal beam of length $l$. The resulting unit cell is a rhombus of side length $\sqrt{3}\, l$ and area $3\sqrt{3}\, l^2/2$.

Assuming that all beams have identical cross-sectional area $A$, moment of inertia $J$, and slenderness $\Lambda$, the elasticity tensor of the equivalent continuum can be expressed in Voigt contracted form as 
\begin{equation}
\label{eq:hexagonal_elastic_lattice_elasticity_tensor}
    \begin{bmatrix}
       \tilde \fE_{1111} &\tilde\fE_{1122} &\tilde\fE_{1112}\\
        \cdot &\tilde\fE_{2222} &\tilde\fE_{2212}\\
        \cdot &\cdot &\tilde\fE_{1212}
    \end{bmatrix} = \frac{\Lambda^2 + 12\,(\phi^2 + 3)}{2 \sqrt{3} \, (\Lambda^2 + 12\,(\phi^2 + 1))}
    \begin{bmatrix}
        1       & \frac{\Lambda^2+12\,(\phi^2-1)}{\Lambda^2 + 12\,(\phi^2 + 3)}    & 0 \\
        \cdot   & 1                                                     & 0 \\ 
        \cdot   & \cdot                                                 & \frac{24}{\Lambda^2+12\,(\phi^2+3)}
    \end{bmatrix},
\end{equation}
where only two of the three components, made dimensionless through Eq.~\eqref{peretta}, are independent, in accordance with the isotropy conditions~\eqref{eq:isotropy_elasticity_components}. The dimensionless Young's modulus and Poisson's ratio are obtained by inverting the elasticity tensor, and are given by
\begin{equation}
\label{eq:hexagonal_equivalent_eng_const}
    \tilde E = \frac{16\sqrt{3}}{\Lambda^2 + 12(\phi^2 + 3)}, \quad
    \nu = \frac{\Lambda^2 + 12(\phi^2 - 1)}{\Lambda^2 + 12(\phi^2 + 3)},
\end{equation}
whereas the shear modulus is $\tilde G = \tilde E/(2(1+\nu))$, as expected for an isotropic linear elastic material.

\begin{figure}[hbt!]
    \centering
    \begin{subfigure}[]{.96\textwidth}
        \centering
        \includegraphics[width=\textwidth]{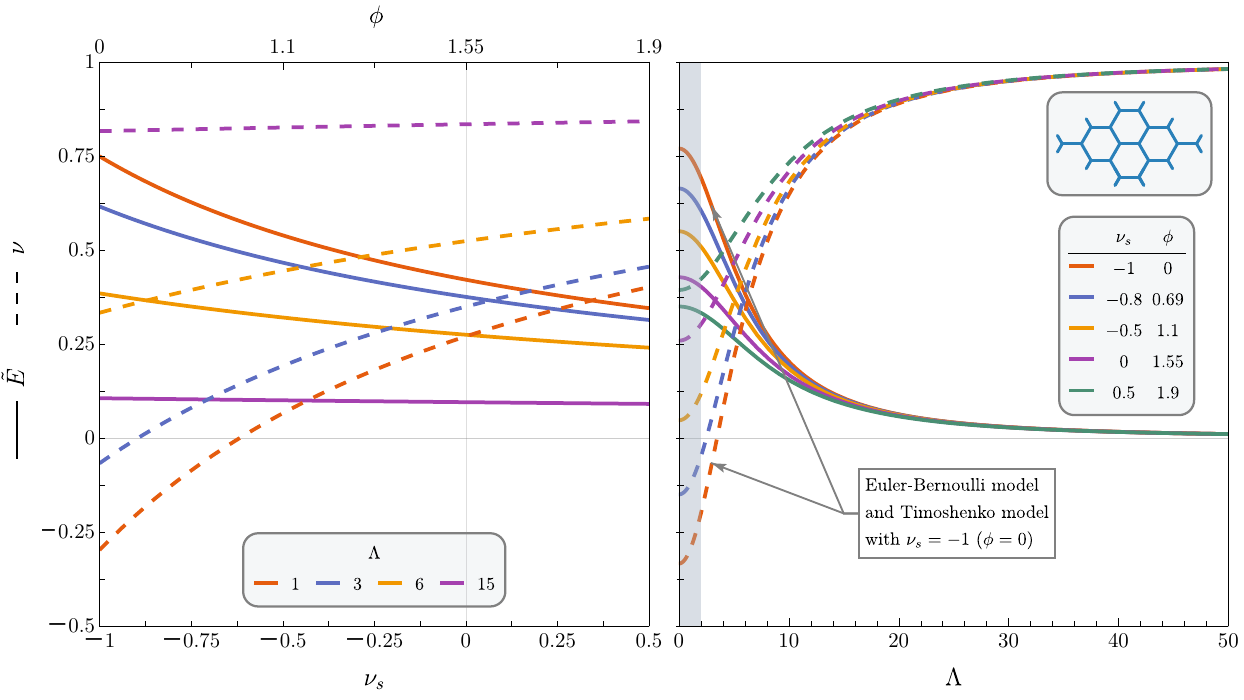}
    \end{subfigure}
    \caption{
    Effective Young's modulus, $\tilde E$, and Poisson's ratio, $\nu$ plotted as functions of the Poisson's ratio $\nu_s$ of the material composing the beams (left), and the slenderness $\Lambda$ (right), for $\kappa = 5/6$. The engineering constants of the effective material can be tuned by properly selecting $\Lambda$ and $\nu_s$.
    }
    \label{fig:hexagonal_elastic_lattice_E_nu}
\end{figure}

Assuming the beams are made of an isotropic material and have a rectangular cross-section, so that $\kappa=5/6$, the Young's modulus $\tilde E$ and Poisson's ratio $\nu$ can be expressed as functions of the slenderness $\Lambda$ and the Poisson's ratio of the base material $\nu_s$, by using relation~\eqref{costantazze2}. Under these assumptions, the engineering constants are shown in Fig.~\ref{fig:hexagonal_elastic_lattice_E_nu} as functions of $\nu_s$ (left) and $\Lambda$ (right). 

It is worth noting that the engineering constants of the material equivalent to the hexagonal grid, given in Eq.~\eqref{eq:hexagonal_equivalent_eng_const}, can be tuned by appropriately selecting $\Lambda$ and $\nu_s$. In particular, the equivalent Young's modulus, $\tilde E$, increases as both the slenderness and the Poisson's ratio of the beams decrease. Conversely, the equivalent Poisson's ratio $\nu$ decreases as $\Lambda$ and $\nu_s$ decrease, reaching its minimum value for $\nu_s=-1$, which also corresponds to the Euler-Bernoulli beam model, and in the limit $\Lambda \to 0$.

The fact that bending moments are essential for the equilibrium of the hexagonal grid under analysis (in contrast to the triangular grid discussed in the previous section) is reflected in the observation that smaller values of $\Lambda$ are required, compared to the triangular lattice, to obtain a sufficiently stiff equivalent continuum. In particular, for $\Lambda > 50$, the effective Young's modulus becomes negligible, rendering the equivalent continuum extremely compliant. Simultaneously, the effective Poisson's ratio approaches its upper limit of $1$ for both the Euler-Bernoulli and Timoshenko models, causing the effective shear modulus $\tilde G$ to vanish.

In fact, for the grid design shown in Fig.~\ref{fig:hexagonal_elastic_lattice_geometry}, bending stiffness is essential. The beams cannot be connected through hinges, as such a configuration would render the grid incapable of sustaining any stress state other than isotropic. This behavior can be demonstrated by considering the limit in which the bending stiffness of the beams vanishes and observing how the equivalent elastic properties change. Since $\tilde E$ and $\tilde G$ depend on the slenderness, and according to Eq.~$\eqref{costantazze1}$, reducing the bending stiffness of the beams corresponds to increasing their slenderness. Consequently, in the limit $\Lambda \to \infty$ one obtains
\begin{equation}
    \tilde E_\infty^{\text{hex}} = \lim_{\Lambda \to \infty} \tilde E^{\text{hex}} = 0, 
    \qquad 
    \tilde G_\infty^{\text{hex}} = \frac{\tilde E_\infty^{\text{hex}}}{2(1+ \nu_\infty^{\text{hex}})} = 0,
    \qquad 
    \nu_\infty^{\text{hex}} = \lim_{\Lambda \to \infty} \nu^{\text{hex}} = 1,
\end{equation}
that is, both the effective Young's and shear moduli vanish, allowing rigid-body displacements, while the Poisson's ratio reaches its 2D upper limit of $1$, as shown in Fig.~\ref{fig:hexagonal_elastic_lattice_E_nu} (right) as slenderness approaches $50$.

Conversely, it can be noted that the Young's and shear moduli for the triangular grid shown in Fig.~\ref{fig:triangular_elastic_lattice_geometry} do not vanish as the bending stiffness approaches zero, but instead converge to the finite values 
\begin{equation}
    \tilde E_\infty^{\text{tri}} = \frac{2}{\sqrt{3}},
    \qquad
    \tilde G_\infty^{\text{tri}} = \frac{\sqrt{3}}{4},
    \qquad
    \nu_\infty^{\text{tri}}  = \frac{1}{3},
\end{equation}
so that the effective Poisson's ratio attains the 2D plane stress value corresponding to the pin-jointed triangular lattice \cite{hutchinson_structural_2006,slepyan_triangular_2001}.

\begin{figure}[hbt!]
    \centering
    \begin{subfigure}[]{.48\textwidth}
        \centering
        \includegraphics[width=\textwidth]{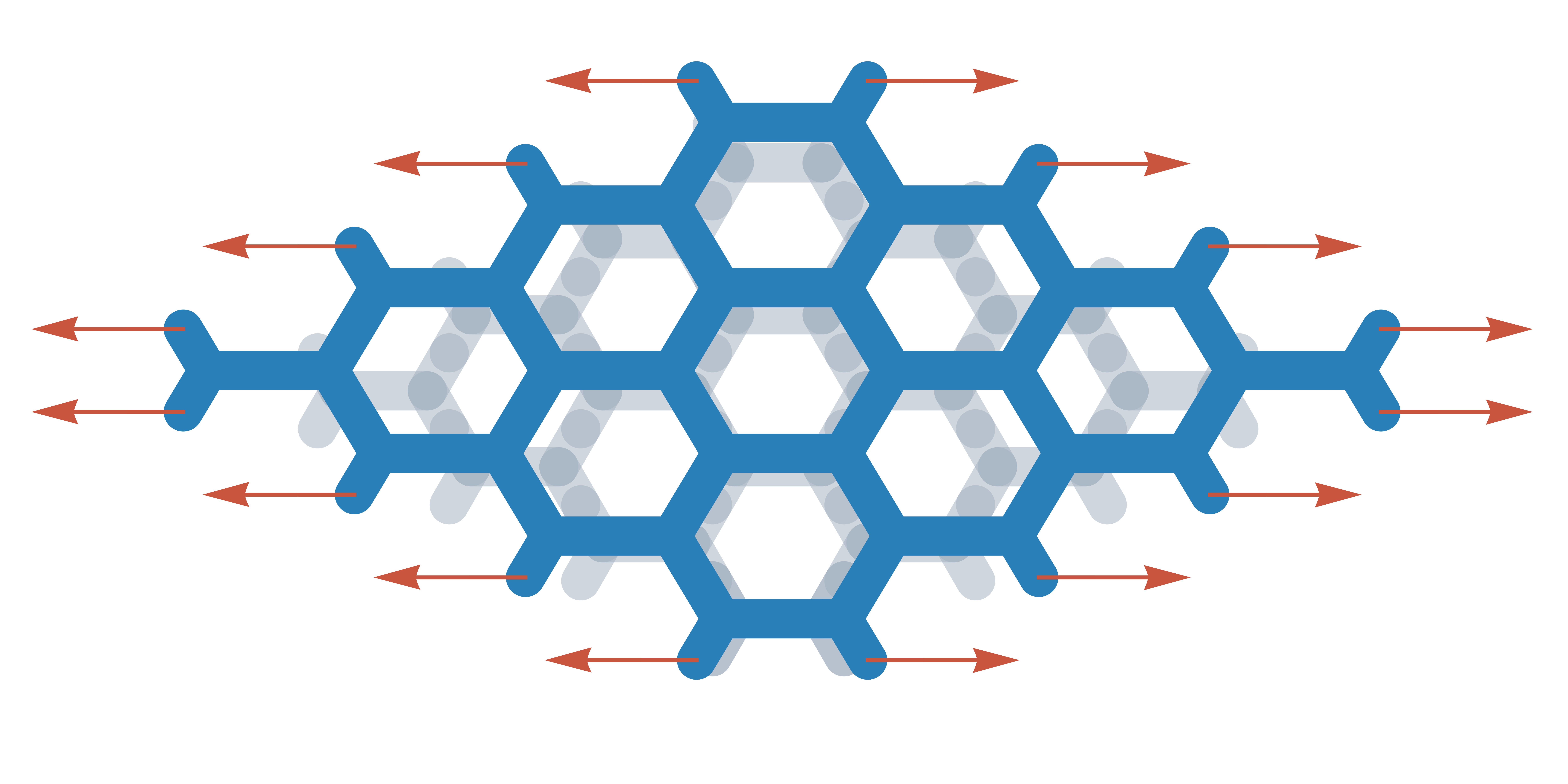}
        \caption{Euler-Bernoulli beam model with $\Lambda=1$, yielding $\nu=-0.297$}
    \end{subfigure}
    \begin{subfigure}[]{.48\textwidth}
        \centering
        \includegraphics[width=\textwidth]{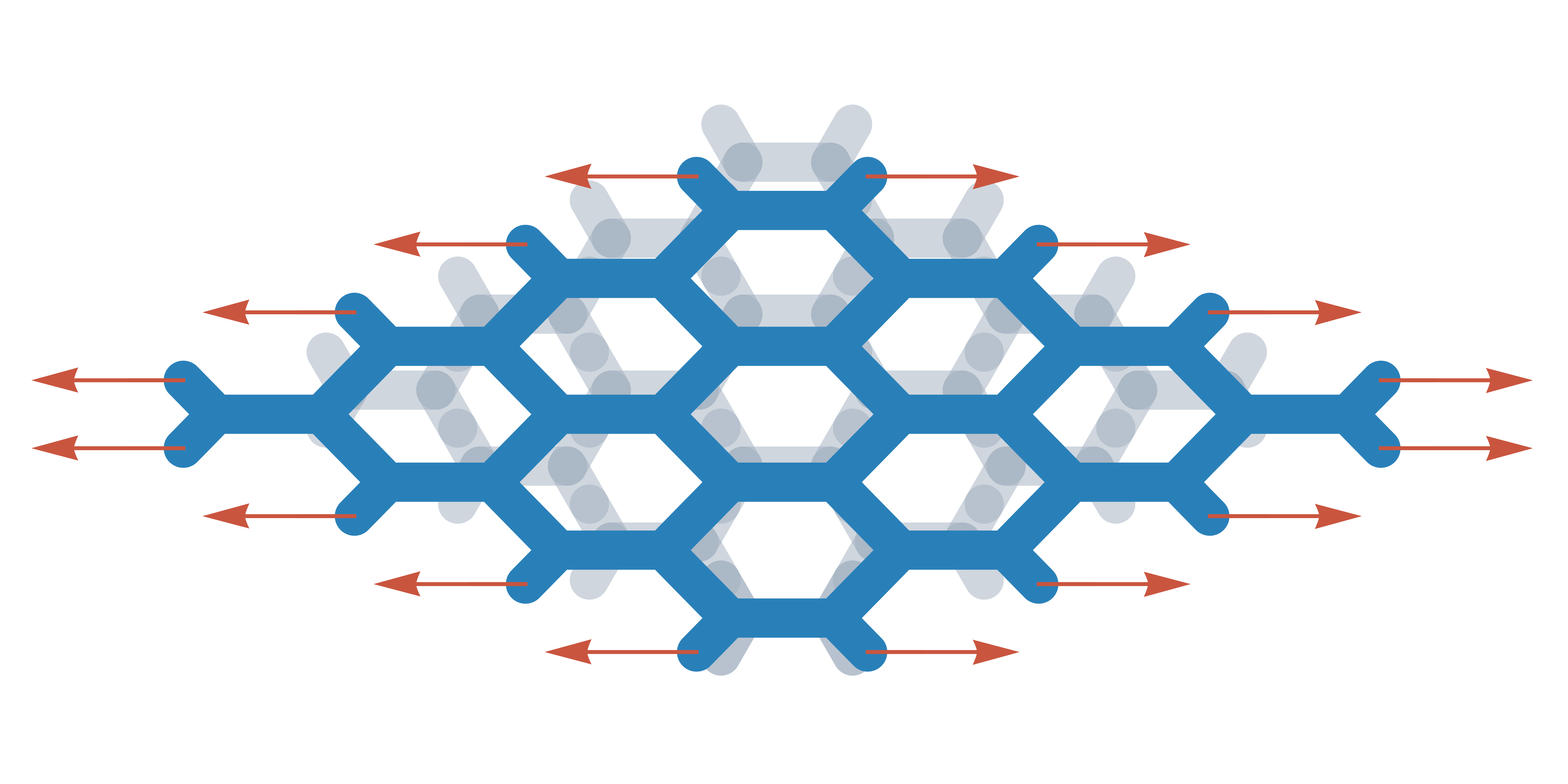}
        \caption{Timoshenko beam model with $\Lambda=1$ and $\nu_s=0.3$, yielding $\nu=0.353$}
    \end{subfigure}
    \caption{
    As in Fig.~\ref{fig:triangular_elastic_deformed}, except that the lattice is now hexagonal. The Timoshenko model (b) corrects the false auxeticity exhibited by the Euler-Bernoulli model (a).
    }
    \label{fig:hexagonal_elastic_lattice_deformed}
\end{figure}

The response of the hexagonal lattice under uniaxial horizontal stress is shown in Fig.~\ref{fig:hexagonal_elastic_lattice_deformed} for stubby beams, illustrating the difference between the Euler-Bernoulli beam model in panel (a) and the Timoshenko beam model with $\nu_s=0.3$ in panel (b). As expected, the Euler-Bernoulli beam model exhibits false auxeticity, causing the lattice to expand along direction $\be_2$ when stretched along direction $\be_1$. In contrast, the Timoshenko beam model correctly predicts contraction along direction $\be_2$, resulting in an effective Poisson's ratio greater than that of the base material, $\nu = 0.353 > \nu_s$.

\subsubsection{Triangular lattice with hexagonal rigid inclusions}
\label{rigid_inclusions}
\begin{figure}[hbt!]
    \centering
    \begin{subfigure}[]{0.6\textwidth}
        \centering
        \includegraphics[width=\textwidth, keepaspectratio]{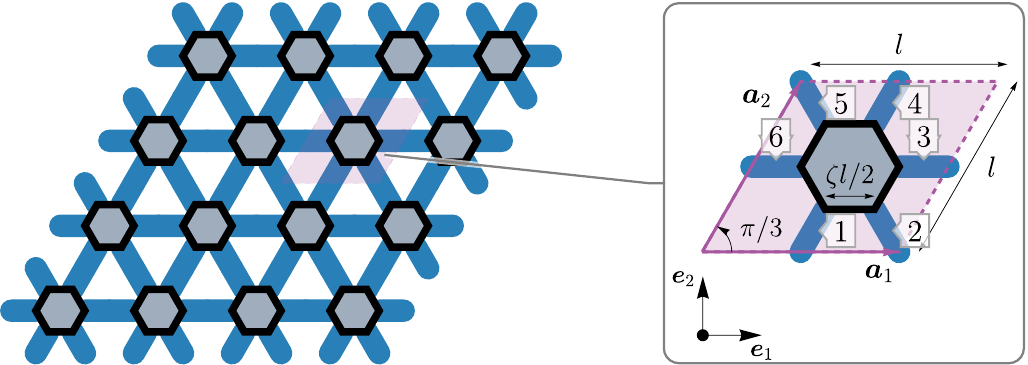}
    \end{subfigure}
    \caption{The unit cell composed of six elastic Timoshenko beams connected to a rigid hexagonal element (right) yields an equilateral triangular lattice, stiffened with rigid inclusions (left).}
    \label{fig:triangular_hexrigid_lattice_geometry}
\end{figure}

The mechanical response of a triangular lattice can be enhanced by embedding rigid hexagonal inclusions, which constrain the deformation of the surrounding beams and increase the overall stiffness, as illustrated in Fig.~\ref{fig:triangular_hexrigid_lattice_geometry}. 

The unit cell has side length $l$ and consists of six identical elastic Timoshenko beams emanating from the vertices of a rigid hexagonal inclusion, as shown in Fig.~\ref{fig:triangular_hexrigid_lattice_geometry} on the right. The length and slenderness of the beams are given by
\begin{equation}
    l_b = l(1-\zeta), \qquad 
    \Lambda = l_b \sqrt{\frac{A}{J}} = l(1-\zeta) \sqrt{\frac{A}{J}},
\end{equation}
where $\zeta \in (0,1)$ is a dimensionless parameter representing the ratio between the inclusion side length and the unit cell side length. The beams are rigidly connected to the hexagonal inclusion, which imposes kinematic constraints at the junctions.

The effective elasticity tensor $\tilde \fE$, whose components have been made dimensionless according to Eq.~\eqref{peretta}, is computed as
\begin{equation}
    \begin{bmatrix}
        \tilde \fE_{1111} & \tilde\fE_{1122} & \tilde\fE_{1112} \\
        \cdot & \tilde\fE_{2222} & \tilde\fE_{2212} \\
        \cdot & \cdot & \tilde\fE_{1212}
    \end{bmatrix} = 
    \frac{3\sqrt{3} \left(\Lambda^2 + 12\phi^2+ 4\right)}{4 (1-\zeta) \left(\Lambda^2 + 12\phi^2\right)} 
    \begin{bmatrix}
        1 & \frac{\Lambda^2 + 12(\phi^2 - 1)}{3\left(\Lambda^2 + 12\phi^2 + 4 \right)} & 0 \\
        \cdot & 1 & 0 \\ 
        \cdot & \cdot & \frac{\Lambda^2 + 12(\phi^2 + 1)}{3(\Lambda^2 + 12\phi^2 + 4)} 
    \end{bmatrix},
\end{equation}
which corresponds to an isotropic solid, since $\tilde \fE_{1111} = \tilde \fE_{2222}$, $\tilde \fE_{1112} = \tilde \fE_{2212} = 0$, and $\tilde \fE_{1212} = (\tilde \fE_{1111} - \tilde \fE_{1122}) / 2$. 

The dimensionless compliance tensor $\tilde \fE^{-1}$ is fully characterized by two independent engineering constants, expressed as functions of the parameters $\zeta$, $\Lambda$, and $\phi$, given by
\begin{equation}
\label{eq:triangular_hexrigid_equivalent_eng_const}
    \tilde E = \frac{2 \left(\Lambda ^2 + 12(\phi^2 + 1)\right)}{\sqrt{3}(1-\zeta) \left(\Lambda^2 + 12 \phi^2 + 4\right)}, 
    \quad
    \nu = \frac{\Lambda^2 + 12 (\phi^2 - 1)}{3 \left(\Lambda^2 + 12 \phi^2 + 4\right)},
\end{equation}
while the effective shear modulus is $\tilde G = \tilde E/(2 (1+\nu))$.

Note that the effective Poisson's ratio $\nu$ is independent of the rigid inclusion size, measured by $\zeta$. Consequently, similarly to the previously analyzed triangular lattice, the Poisson's ratio of the equivalent material lies within the range $-1 < \nu < 1/3$. 

\begin{figure}[hbt!]
    \centering
    \begin{subfigure}[]{.4\textwidth}
        \centering
        \includegraphics[width=\textwidth]{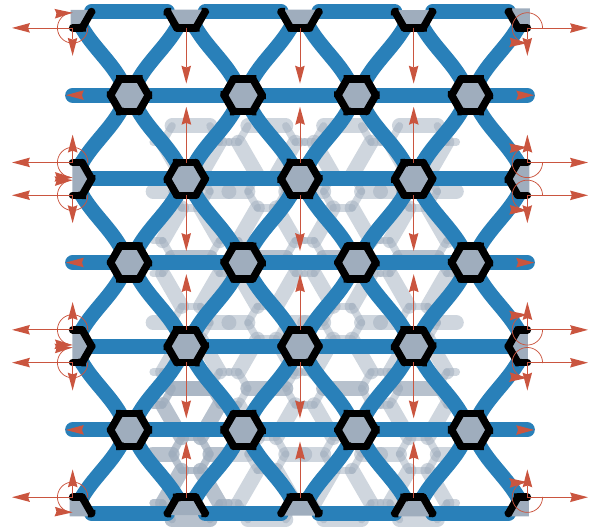}
        \caption{Euler-Bernoulli beam model with $\Lambda=1$, yielding $\nu=-11/15 \approx -0.733$}
    \end{subfigure}
    \begin{subfigure}[]{.4\textwidth}
        \centering
        \includegraphics[width=\textwidth]{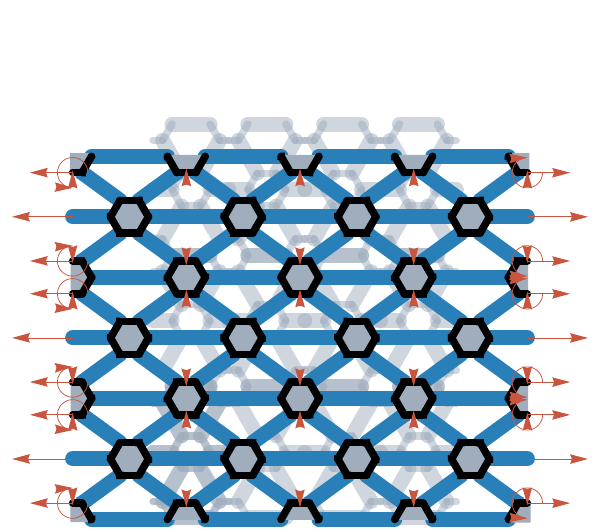}
        \caption{Timoshenko beam model with $\Lambda=1$ and $\nu_s=0.3$, yielding $\nu=0.207$}
    \end{subfigure}
    \caption{
    As in Fig.~\ref{fig:triangular_elastic_deformed}, but for a hexagonal lattice with rigid inclusions ($\zeta=1/2$). The Timoshenko model (b) corrects the false auxetic behavior shown by the Euler-Bernoulli model (a). 
    }
\label{fig:triangular_hexrigid_lattice_deformed}
\end{figure}

Compared to the triangular lattice response shown in Fig.~\ref{fig:triangular_elastic_deformed}, the presence of rigid inclusions results in a stiffer response under uniaxial horizontal uniform stress, as illustrated in Fig.~\ref{fig:triangular_hexrigid_lattice_deformed}. While the inclusions reduce the overall displacement magnitudes, the ratio between axial and transverse deformations remains unchanged for the two grids, as evidenced by a comparison of Figs.~\ref{fig:triangular_elastic_deformed} and \ref{fig:triangular_hexrigid_lattice_deformed}.

\subsection{Re-entrant lattice and true auxeticity}
Two examples of re-entrant grids composed of Timoshenko beams are examined below, as these geometries are well-known to produce auxetic response in the equivalent material \cite{evans_auxetic_1991,evans_molecular_1991,gibson_cellular_1997,lakes_foam_1987,lakes_advances_1993,masters_models_1996}. 
In these lattices, flexural stiffness plays a crucial role, similarly to the hexagonal geometry discussed in Section~\ref{sec:hexagonal_lattice}, so that taking the limit $\Lambda \to \infty$ results in a collapse mechanism for the re-entrant lattice, corresponding to a singular constitutive tensor for the effective material.

To maintain generality, the shear stiffness will be incorporated through the parameter $\phi$, as defined in Eqs.~$\eqref{costantazze2}$ and $\eqref{eq:lambda-phi_chain}$. This parameter ranges from zero (corresponding to the Euler-Bernoulli beam model or to the Timoshenko model with $\nu_s=-1$) to infinity (representing a beam with extremely high shear deformability, such as the microstructured beam illustrated in Fig.~\ref{trave_a_taglio}).
Accordingly, the Poisson's ratio of the beam constituent material $\nu_s$ will be replaced by the shear coefficient $\phi$. The particular case of beams made from an isotropic material with $\nu_s = 0.3$ and a rectangular cross-section ($\kappa=5/6$) corresponds to $\phi=1.76$.

\begin{figure}[hbt!]
    \centering
    \begin{subfigure}[]{0.6\textwidth}
        \centering
        \includegraphics[width=\textwidth, keepaspectratio]{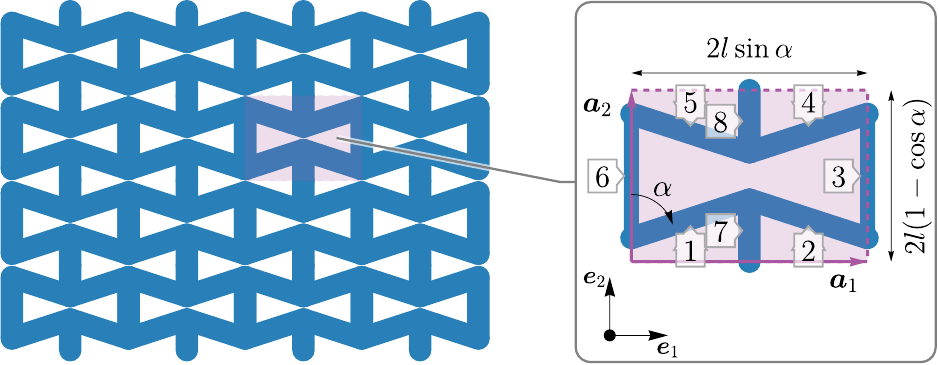}
    \end{subfigure}
    \caption{
    A lattice (left) composed of re-entrant unit cells (right), each consisting of eight Timoshenko beams.
    }
    \label{fig:re-entrant_hexagonal_lattice_geometry}
\end{figure}

The first example concerns a re-entrant hexagonal unit cell, shown in Fig.~\ref{fig:re-entrant_hexagonal_lattice_geometry}, consisting of two vertically aligned, \lq arrow-shaped' beams connected by two additional vertical beams along the lateral edges of the cell. The geometry of the unit cell is governed by the re-entrant angle $\alpha$, which is restricted to the interval $(\pi/3, \pi/2)$. At the upper bound $\alpha = \pi/2$, the internal cavity assumes a rectangular shape. For values of $\alpha > \pi/2$, the configuration ceases to be re-entrant, and the auxetic behavior is lost. At the lower bound $\alpha = \pi/3$, the arrow tips make contact, rendering further reduction of $\alpha$ geometrically infeasible due to beam interference—unless the lattice joints are designed to allow the beams to operate on separate planes.

All beams are assumed to be identical, with length $l$, cross-sectional area $A$, and slenderness $\Lambda$. Although the side lengths of the unit cell depend on $l$ and the re-entrant angle $\alpha$, the unit cell remains rectangular for all values of $(l,\alpha)$, and the two vectors $\ba_1$, $\ba_2$, defining the direct basis, remain mutually orthogonal. The lattice is generated by periodically repeating the unit cell along the directions $\ba_1$, $\ba_2$, resulting in the geometry shown in Fig.~\ref{fig:re-entrant_hexagonal_lattice_geometry}. 

According to Eq.~\eqref{peretta}, the dimensionless elasticity tensor can be expressed in Voigt form as 
\begin{multline}
\label{eq:re-entrant_hexagonal_lattice_elasticity_tensor}
    \begin{bmatrix}
        \tilde \fE_{1111} & \tilde\fE_{1122} & \tilde\fE_{1112} \\
        \cdot & \tilde\fE_{2222} & \tilde\fE_{2212} \\
        \cdot & \cdot & \tilde\fE_{1212}
    \end{bmatrix} =
    \\[3mm]
    = \dfrac{k_0 \sin\alpha}{2} 
    \begin{bmatrix}
        \dfrac{(\Theta - 12) (1 - \cos 2\alpha) + 72}{\sin\alpha\, \tan\frac{\alpha}{2}}
        & -2 (\Theta - 12) \cos\alpha  
        & 0
        \\
        \cdot 
        & \dfrac{(\Theta - 12) (1 + \cos 2\alpha) + 24}{\sin\alpha\,\cot\frac{\alpha}{2}}
        & 0
        \\
        \cdot
        & \cdot
        & \dfrac{8k_1}{k_0}
    \end{bmatrix},
\end{multline}
where the coefficients $k_0$ and $k_1$ are defined as
\begin{equation}
\label{eq:k0-k1}
    k_0 = \frac{1}{\Theta(2 + \cos 2\alpha) + 12(1-\cos 2\alpha)}, 
    \quad 
    k_1 = \frac{1}{\Theta(1 + \cos\alpha) + 4(1-\cos\alpha)},
\end{equation}
both depending on the parameter $\Theta$, that combines $\Lambda$ and $\phi$ as
\begin{equation}
\label{eq:Theta}
    \Theta = \Lambda^2 + 12\phi^2.
\end{equation}

The analytical expressions for the dimensionless engineering constants, corresponding to the re-entrant hexagonal lattice, are 
\begin{equation}
\begin{aligned}
    & \tilde E_1 = \dfrac{12 \cot\frac{\alpha}{2}}{(\Theta + 12 \tan^2\alpha) \cos^2\alpha},
    \quad
    \tilde E_2 = \dfrac{12 \tan\frac{\alpha}{2}}{(\Theta + 12 \cot^2\alpha) \sin^2\alpha + 24},
    \quad
    \tilde G_{12} = 4k_1 \sin\alpha, \\
    & \nu_{12} = -\dfrac{\Theta - 12}{\Theta + 12 \tan^2\alpha}(1+\sec\alpha),
    \quad
    \nu_{21} = -2\dfrac{\Theta - 12}{(\Theta + 12 \cot^2\alpha) \sin^2\alpha + 24} \cos\alpha\, \sin^2\frac{\alpha}{2}.
\end{aligned}
\end{equation}
Tensor~\eqref{eq:re-entrant_hexagonal_lattice_elasticity_tensor} satisfies the condition $\fE_{1112} = \fE_{2212} = 0$, implying that the effective material exhibits orthotropic symmetry. This symmetry class is characterized by \textit{four} independent elasticity components, since the two Poisson's ratios, $\nu_{12}$ and $\nu_{21}$, are related through the symmetry of the compliance tensor $\fE^{-1}$
\begin{equation}
    \frac{\nu_{12}}{E_1} = \dfrac{\nu_{21}}{E_2}.
\end{equation}

\begin{figure}[hbt!]
    \centering
    \begin{subfigure}[]{.49\textwidth}
        \centering
        \includegraphics[width=\textwidth]{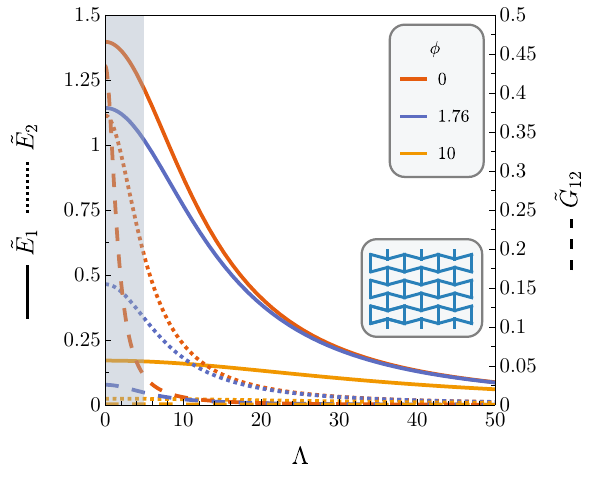}
    \end{subfigure}
    \begin{subfigure}[]{.49\textwidth}
        \centering
        \includegraphics[width=\textwidth]{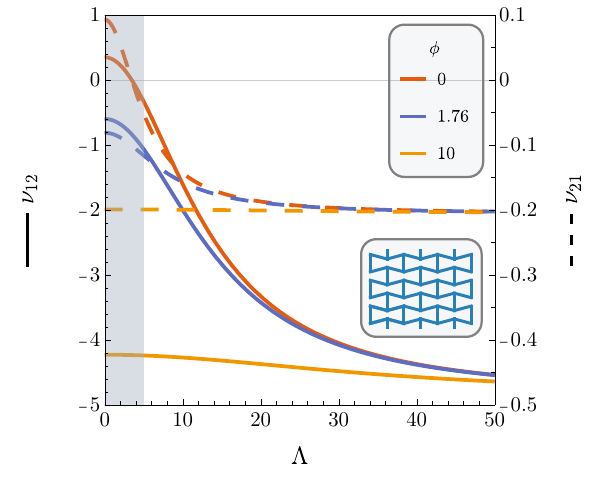}
    \end{subfigure}
    \caption{
    Effective Young's and shear moduli (left), and Poisson's ratios (right) plotted as functions of the slenderness $\Lambda$, for a fixed re-entrant angle $\alpha=75^\circ$. Three values of shear parameter $\phi$ are considered: Euler-Bernoulli theory (red lines), Timoshenko theory for an isotropic material with a rectangular cross-section (blue lines), and Timoshenko theory for a highly shear-deformable material (yellow lines). Note that there are two different scales for the elastic properties on the left and the right of the plots.
    }
    \label{fig:re-entrant_elastic_lattice_eng_const_lambda}
\end{figure}

The engineering constants are plotted in Fig.~\ref{fig:re-entrant_elastic_lattice_eng_const_lambda} as functions of the slenderness $\Lambda$, for both the Euler-Bernoulli beam theory (red lines) and the Timoshenko beam theory, with a fixed re-entrant angle $\alpha=75^\circ$. 

The shear deformability of the beams reduces the stiffness of the effective material compared to the Euler-Bernoulli beam model, as shown in Fig.~\ref{fig:re-entrant_elastic_lattice_eng_const_lambda} (left), with the effect becoming more pronounced as beams become stubby. This reduction is moderate for $\tilde E_1$ (with values $\tilde E_1 = 1.36$ and $\tilde E_1 = 1.12$ for $\phi = 0$ and $\phi = 1.76$, respectively), but more significant for $\tilde E_2$ (decreasing from $\tilde E_2 = 0.32$ to $\tilde E_2 = 0.15$) and the shear modulus $\tilde G_{12}$ (decreasing from $\tilde G_{12} = 0.48$ to $\tilde G_{12} = 0.07$), for $\Lambda=2$, $\alpha=75^\circ$. However, the influence of shear deformability diminishes rapidly as the beams become slender, with all curves converging to the same limiting values ($\tilde E_1 = 0.09$, $\tilde E_2 = 0.004$, and $\tilde G_{12} = 0.001$ at $\Lambda=50$ and $\alpha=75^\circ$) and vanishing in the limit $\Lambda \to \infty$. In other words, for the re-entrant cell geometry, compared to the previously analyzed grids, shear deformability leads to a significant reduction in the elastic properties of the equivalent material when the beams are stubby.

It is worth noting that, for the chosen re-entrant angle $\alpha=75^\circ$, auxetic behavior is observed for all values of $\phi$, except in the range $\Lambda < 2\sqrt{3}$, where the Euler-Bernoulli model predicts positive Poisson's ratios. However, when $\Lambda \lesssim 4$, geometric interference occurs between the beams. In this regime, the modeling approaches described in Section~\ref{sec:decoupled_plane_stress} must be adopted to avoid physical interpenetration and to preserve model consistency.

\begin{figure}[hbt!]
    \centering
    \begin{subfigure}[]{.48\textwidth}
        \centering
        \includegraphics[width=\textwidth]{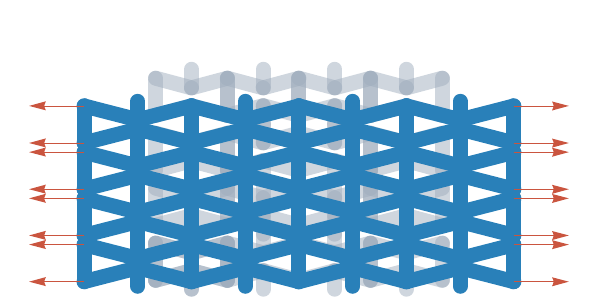}
        \caption{Euler-Bernoulli beam model with $\Lambda=1$, $\alpha = 75^\circ$, yielding $\nu_{12}=0.318$}
    \end{subfigure}
    \begin{subfigure}[]{.48\textwidth}
        \centering
        \includegraphics[width=\textwidth]{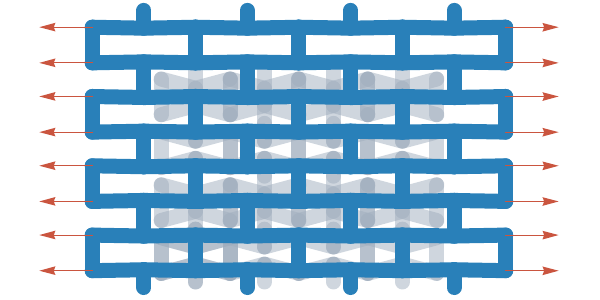}
        \caption{Timoshenko beam model with $\Lambda=1$, $\phi=1.76$, $\alpha = 75^\circ$, yielding $\nu_{12}=-0.62$}
    \end{subfigure}
    \caption{
    As in Fig.~\ref{fig:triangular_elastic_deformed}, except that the lattice is now re-entrant ($\alpha=75^\circ$), the Timoshenko model (b) correctly exhibits auxetic behavior, in contrast to the Euler-Bernoulli model (a). 
    }
    \label{fig:re-entrant_lattice_deformed}
\end{figure}

The mechanical response of the re-entrant lattice under horizontal uniaxial stress is shown in Fig.~\ref{fig:re-entrant_lattice_deformed}. The beam slenderness is $\Lambda=1$, a low value chosen to emphasize the effect. The lattice on the left, constructed with Euler-Bernoulli beams, exhibits an incorrect non-auxetic response, while the lattice on the right, modeled with Timoshenko beams, correctly displays auxetic behavior.

\begin{figure}[hbt!]
    \centering
    \begin{subfigure}[]{0.6\textwidth}
        \centering
        \includegraphics[width=\textwidth, keepaspectratio]{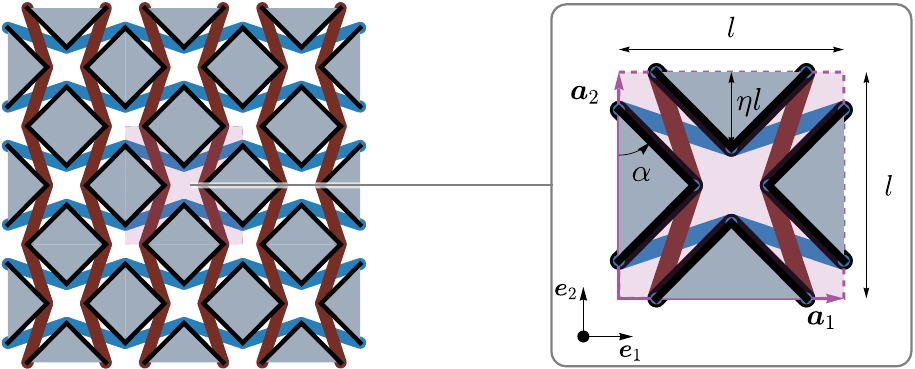}
    \end{subfigure}
    \caption{
    The re-entrant lattice with square rigid inclusions (left) is formed by tassellating the square unit cell (right) along the directions $\be_1, \be_2$. The rigid inclusions (gray areas surrounded by black thick lines) are fully characterized by the parameters $\eta$ and $\alpha$ that satisfy condition~\eqref{eq:geometric_condition}.
    }
    \label{fig:re-entrant_sqrigid_lattice_geometry}
\end{figure}

The second example considers a re-entrant grid of Timoshenko beams incorporating square rigid inclusions. The unit cell, illustrated in Fig.~\ref{fig:re-entrant_sqrigid_lattice_geometry} (right), consists of four rigid inclusions connected at their corners by eight Timoshenko beams. It is important to note that the beams shown in red and those shown in blue in the figure are positioned on separate planes and connected only through the rigid inclusions. The symmetry of the unit cell ensures that the equivalent continuum behaves as a cubic material. As a result, the components of the elasticity tensor satisfy the following conditions
\begin{equation}
\label{eq:cubic_symmetry_E}
    \fE_{1111} = \fE_{2222}, \quad \fE_{1112} = \fE_{2212} = 0.
\end{equation}

\begin{figure}[hbt!]
    \centering
    \begin{subfigure}[]{.96\textwidth}
        \centering
        \includegraphics[height=.3\textwidth]{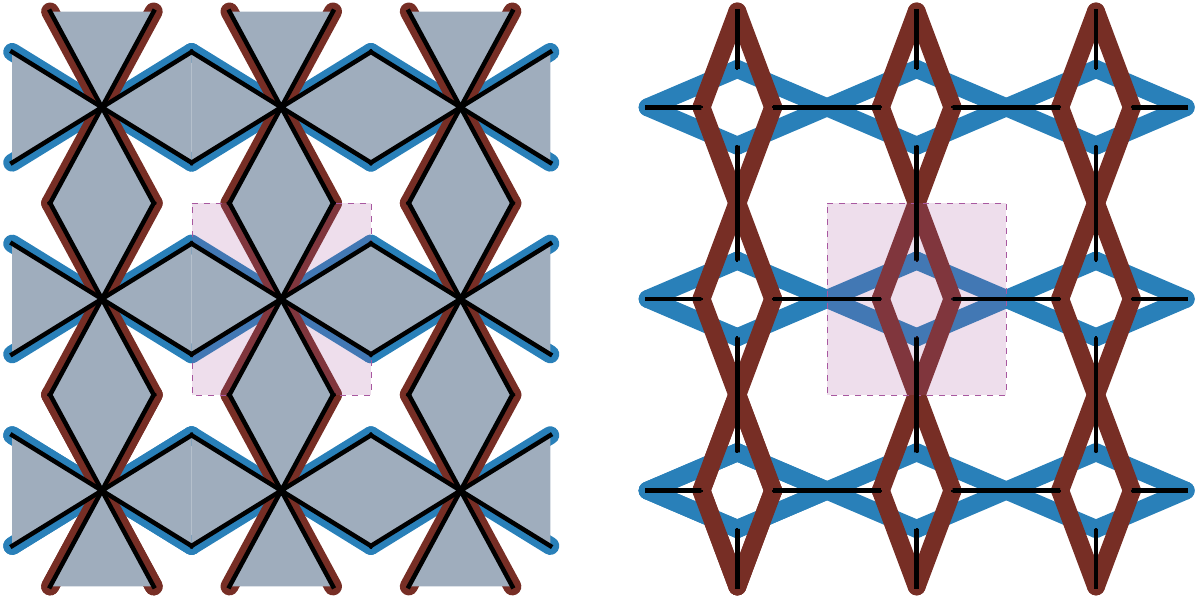}
    \end{subfigure}
    \caption{
    Limit geometries of the re-entrant square lattice with rigid inclusions. Elastic beams overlap the boundary of the rigid inclusions for $\alpha=\pi/3$ and $\eta = 1/2$ (left); rigid inclusions degenerate into rigid rods for $\alpha=\pi/2$ and $\eta=0.3$ (right).
    }
    \label{fig:re-entrant_sqrigid_lattice_limits_compare}
\end{figure}

The length of the elastic beams can be expressed as a function of $\alpha$ and $\eta$ by
\begin{equation*}
    l_b = \frac{l}{2} \sqrt{(2\eta\cot\alpha + 2\eta - 1)^2 + 1},
\end{equation*}
where $l$ is the dimension of the unit cell, as illustrated in Fig.~\ref{fig:re-entrant_sqrigid_lattice_geometry} (right). The parameters $\eta$ and $\alpha$ must satisfy the geometric conditions
\begin{equation}
\label{eq:geometric_condition}
   0 \leq \eta \leq \frac{1}{2}, \qquad 0 \leq \alpha \leq \pi, \qquad -\frac{1}{2} \leq \eta\cot\alpha \leq \frac{1}{2},
\end{equation}
Consequently, the slenderness of the beams is given by
\begin{equation}
    \Lambda = l_b \sqrt{\frac{A}{J}} = \frac{l}{2}  \sqrt{\frac{A}{J}} \sqrt{(2\eta\cot\alpha + 2\eta - 1)^2 + 1}.
\end{equation}
Two interesting limiting cases occur: when $\alpha=\pi/3$ and $\eta=1/2$, the rigid inclusions intersect at the center of the unit cell; and when $\alpha=\pi/2$ and $\eta=0.3$, the rigid inclusions flatten, effectively becoming rigid rods. These limiting configurations are shown in Fig.~\ref{fig:re-entrant_sqrigid_lattice_limits_compare}.

Keeping in mind conditions~\eqref{eq:cubic_symmetry_E} and using the dimensionless variables defined with Eq.~\eqref{peretta}, the expressions for the non-zero elastic constants are 
\begin{equation}
    \begin{aligned}
        &\tilde \fE_{1111} = 
        \dfrac{\sqrt{2}(\Theta+12)}{\Theta\sqrt{1+2\eta(1+\cot\alpha)[\eta(1+\cot\alpha)-1]}}, \\[3mm]
        &\tilde \fE_{1212} = 
        \frac{4k_3\sqrt{k_2^2+1}(k_2-\cot\alpha)^2}{(1+\cot\alpha)^2 \{\Lambda^2(k_2+1)^2(12k_2^2+\Theta)(1-\sin 2\alpha) + 4\Theta (k_2^2+1)^2(1+\sin 2\alpha)\}} \\[3mm]
        &\tilde \fE_{1122} = 
        -\dfrac{\sqrt{2}(\Theta-12)[2\eta(1+\cot\alpha)-1]}{\Theta\{1+2\eta(1+\cot\alpha)[\eta(1+\cot\alpha)-1]\}^{3/2}},
    \end{aligned}
\end{equation}
which lead to the following three dimensionless engineering constants
\begin{equation}
\label{eq:re-entrant_sqrigid_E_G_nu}
    \begin{aligned}
        &\tilde E = \frac{\sqrt{2}}{4}\frac{(\Theta+12)^2(k_2^2 + 1)^2 - 4k_2^2 (\Theta-12)^2}{\Theta(\Theta+12)\Big[(k_2^2+1)/2\Big]^{5/2}},
        \quad
        \tilde G = \tilde E_{1212},
        \quad
        \nu = - \frac{2k_2(\Theta-12)}{(\Theta+12)(k_2^2+1)},
    \end{aligned}
\end{equation}
where $\Theta$ is defined in Eq.~\eqref{eq:Theta}, while 
\begin{equation}
    k_2 = \sqrt{4\frac{l_b^2}{l^2}-1},
    \quad 
    k_3 = \Bigl(12(k_2+1)^2 + \Theta(k_2-1)^2\Bigr)(1+\sin 2\alpha) + 3\Lambda^2(k_2+1)^2(1-\sin 2\alpha),
\end{equation}
depend on the geometry of the unit cell and on Poisson's ratio of the beam material, included in the parameter $\phi$.

\begin{figure}[hbt!]
    \centering
    \begin{subfigure}[]{.96\textwidth}
        \centering
        \includegraphics[width=\textwidth]{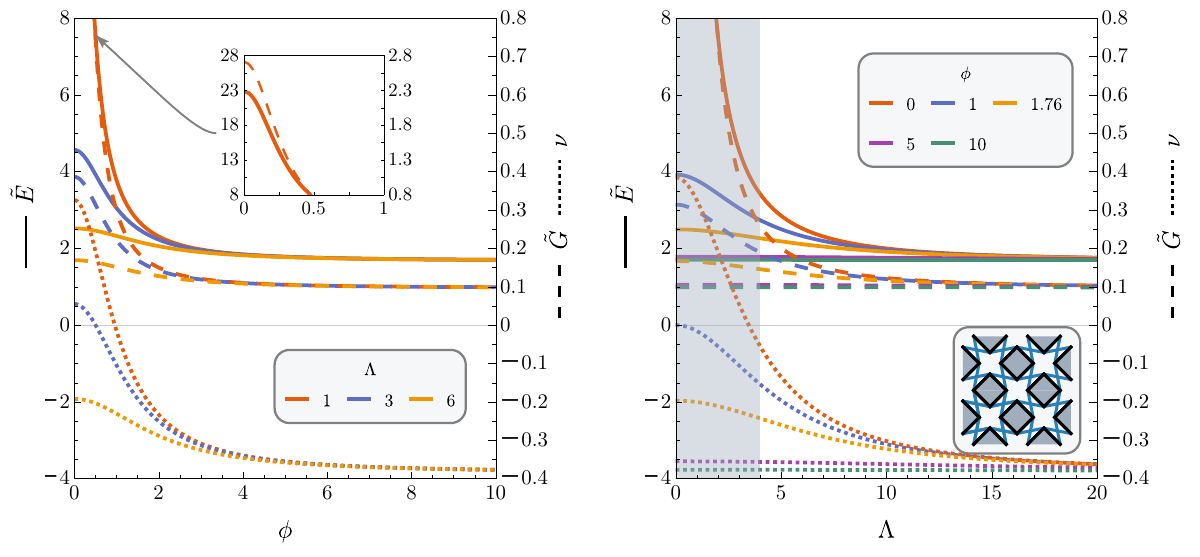}
    \end{subfigure}
    \caption{
    Effective Young's modulus, shear modulus, and Poisson's ratio as functions of the shear coefficient $\phi$ (left) and slenderness $\Lambda$ (right), for fixed geometric parameters $\eta = 0.3$ and $\alpha = \pi/4$. Note the two different scales: one for $\tilde E$ on the left, and another for $\tilde G$ and $\nu$ on the right of the frames.
    }
    \label{fig:re-entrant_sqrigid_lattice_E_G_nu}
\end{figure}

The elastic constants given in Eqs.~\eqref{eq:re-entrant_sqrigid_E_G_nu} are plotted in Fig.~\ref{fig:re-entrant_sqrigid_lattice_E_G_nu} as functions of $\phi$ (left) and $\Lambda$ (right), for $\eta = 0.3$ and $\alpha = \pi/4$, which satisfy condition~\eqref{eq:geometric_condition}. The figure shows that, compared to the Euler-Bernoulli model, the Timoshenko beam model reduces the elastic properties of the effective material, particularly at low slenderness. It can also be observed that all the engineering constants defined in Eq.~\eqref{eq:re-entrant_sqrigid_E_G_nu} decrease as $\phi$ and $\Lambda$ increase, and eventually approach constant values as $\{\phi, \Lambda\} \to \infty$, Fig.~\ref{fig:re-entrant_sqrigid_lattice_E_G_nu}. However, in the left plot of this figure, the equivalent Poisson's ratio remains positive for $0 < \phi < 1$ even with the Timoshenko beam model, when the slenderness is reasonably small (blue and red dotted lines).

Unlike the case of the triangular lattice, bending stiffness is essential for the grid shown in Fig.~\ref{fig:re-entrant_sqrigid_lattice_geometry}. In the limit of vanishing bending stiffness, the following expressions are obtained
\begin{equation}
    \tilde E_{\infty} = \frac{2(k_2^2 - 1)^2}{(k_2^2 + 1)^{5/2}},
    \qquad 
    \tilde G_{\infty} = 0,
    \qquad 
    \nu_\infty = -\frac{2 k_2}{k_2^2+1},
\end{equation}
showing that the shear modulus $\tilde G_\infty$ vanishes, while $\tilde E_{\infty} > 0$, and $\nu_\infty < 0$ for any combination of $\eta$ and $\alpha$ satisfying the geometric condition~\eqref{eq:geometric_condition}.

\begin{figure}[hbt!]
    \centering
    \begin{subfigure}[]{.96\textwidth}
        \centering
        \includegraphics[width=\textwidth]{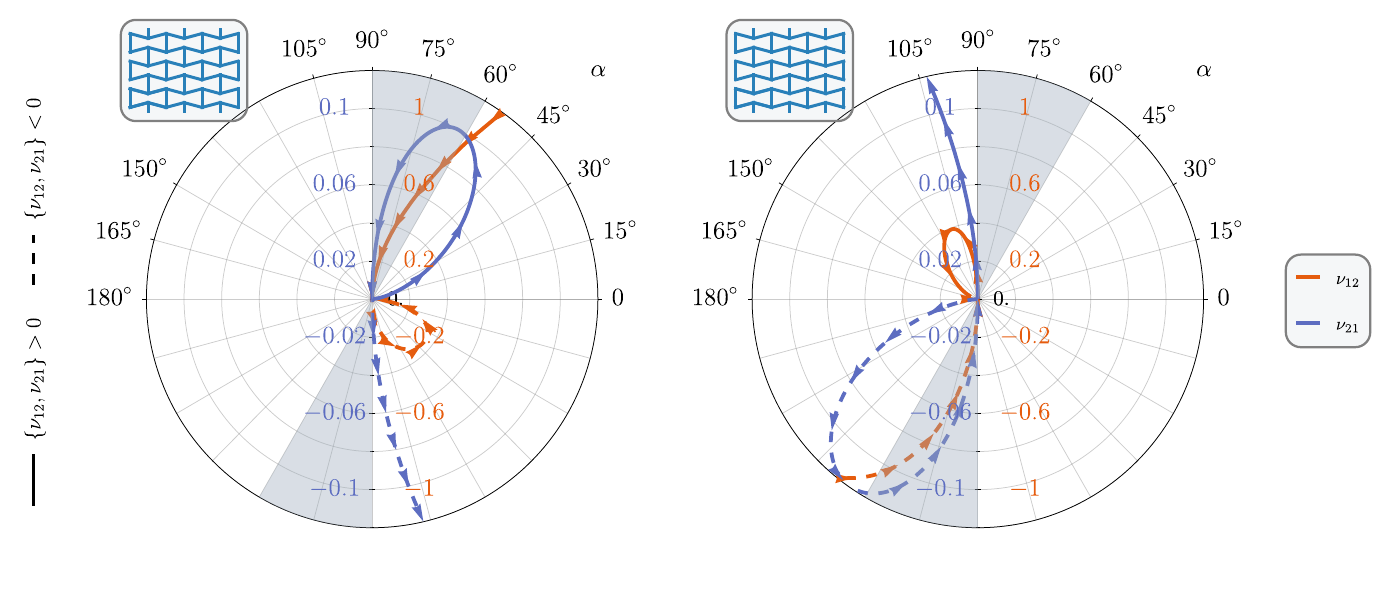}
    \end{subfigure}
    \caption{
    Polar plots of the effective Poisson's ratios for the re-entrant lattice without rigid inclusions, shown as a function of the re-entrant angle $\alpha \in (0, 180^\circ)$: Euler-Bernoulli model, $\phi=0$, (left) and Timoshenko model, $\phi=1.76$, (right). Extremely stubby beams are considered with slenderness $\Lambda = 1$. The gray region marks the range of angles corresponding to re-entrant geometries ($\alpha < 90^\circ$) and no interference between beams ($\alpha > 60^\circ$). Solid curves indicate positive Poisson's ratios, while dashed curves denote negative (auxetic) values. The Timoshenko model (right) predicts consistent auxetic behavior throughout the re-entrant regime, whereas the Euler-Bernoulli model (left) incorrectly suggests non-auxetic behavior in the same range.
    }
    \label{fig:re-entrant_lattice_nu_polar_plot}
\end{figure}

\begin{figure}[hbt!]
    \centering
    \begin{subfigure}[]{.96\textwidth}
        \centering
        \includegraphics[width=\textwidth]{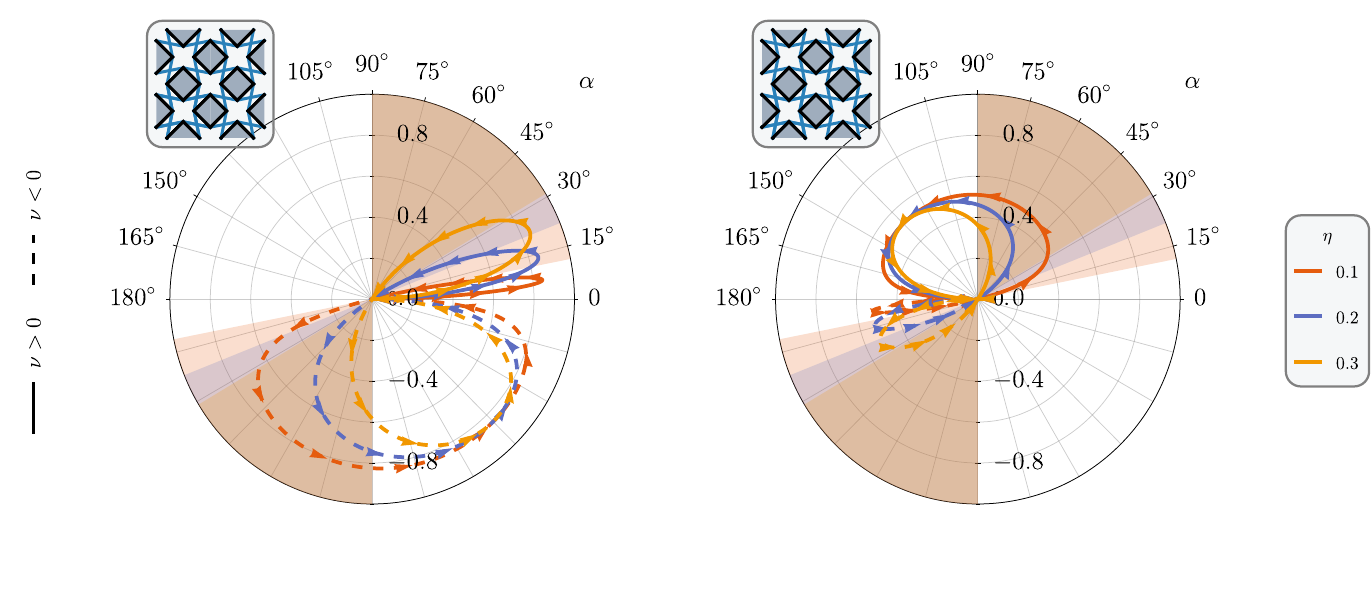}
    \end{subfigure}
    \caption{
    As in Fig.~\ref{fig:re-entrant_lattice_nu_polar_plot}, but for the re-entrant lattice with rigid inclusions. The effective Poisson's ratio $\nu$ is shown for three different values of the geometric parameter $\eta$. Colored regions denote the angular intervals that satisfy both the re-entrant condition $\alpha < 90^\circ$ and the geometric compatibility conditions~\eqref{eq:geometric_condition}. The ranges of angles $\alpha$ for each value of $\eta$ are reported in Table~\ref{tab:eta_alpha}.
    }
    \label{fig:re-entrant_sqrigid_lattice_nu_polar_plot}
\end{figure}

The two re-entrant grids shown in Figs.~\ref{fig:re-entrant_hexagonal_lattice_geometry} and \ref{fig:re-entrant_sqrigid_lattice_geometry} are compared in the polar plots of Figs.~\ref{fig:re-entrant_lattice_nu_polar_plot} and \ref{fig:re-entrant_sqrigid_lattice_nu_polar_plot}, respectively. In these figures, the effective Poisson's ratios are plotted as functions of the re-entrant angle $\alpha$ for both the Euler-Bernoulli model (left panel) (equivalent to the Timoshenko model with $\nu_s = -1$), and the Timoshenko model with $\nu_s = 0.3$ and $\kappa = 5/6$ (right panel). 

In particular, since the equivalent material of the grid shown in  Fig.~\ref{fig:re-entrant_hexagonal_lattice_geometry} is anisotropic, the Poisson's ratios can become unbounded and arbitrarily vary over the real numbers \cite{ting_poisson_ratio_anisotropic_2005, boulanger_poisson_orthorombic_1998}. This behavior is clearly illustrated in Fig.~\ref{fig:re-entrant_lattice_nu_polar_plot}, where $\nu_{12}$ and $\nu_{21}$ are plotted as functions of $\alpha$ for a highly stubby beam ($\Lambda = 1$). For this slenderness value, the Euler-Bernoulli beam theory (left) predicts an unexpected non-auxetic response. In contrast, when shear deformability is taken into account (right), the equivalent material exhibits auxetic behavior for any re-entrant angle $\alpha < 90^\circ$. 

When the re-entrant mechanism is provided by the square rigid inclusions (resulting in a cubic symmetry, Fig.~\ref{fig:re-entrant_sqrigid_lattice_geometry}), the Poisson's ratio remains bounded within the range $(-1,1)$ and the auxetic effect becomes strongly dependent on the geometric parameters, $\alpha$ and $\eta$. In particular, the colored zones in Fig.~\ref{fig:re-entrant_sqrigid_lattice_nu_polar_plot} represent the combinations of $(\alpha, \eta)$ that satisfy the geometric compatibility conditions~\eqref{eq:geometric_condition}. Consequently, outside these regions, the solution is not physically meaningful. Table~\ref{tab:eta_alpha} reports the corresponding ranges of the angle $\alpha$ for the re-entrant lattice with square rigid inclusions at various values of $\eta$. 

\begin{table}[hbt!]
    \centering
    \caption{
    Range of compatible angles $\alpha$ for varying semi-diagonal length of the inclusion, $\eta$, according to Eq.~\eqref{eq:geometric_condition}. Note that the upper limit, $\alpha_{\text{max}}=90^\circ$ is independent of $\eta$ and corresponds to the case where the square inclusions collapse into lines (Fig.~\ref{fig:re-entrant_sqrigid_lattice_limits_compare}, right). Additionally, when $\eta=0.5$, the rigid inclusions intersect at the center of the unit cell, causing the elastic beams to overlap the boundaries of the rigid inclusions (see Fig.~\ref{fig:re-entrant_sqrigid_lattice_limits_compare}, left).
    }
    \label{tab:eta_alpha}
    \begin{tabular}{@{}lcc@{}}
    \toprule
    $\eta$ & $\alpha_{\text{min}}~[^\circ]$ & $\alpha_{\text{max}}~[^\circ]$ \\ \midrule
    0.1 & 11.31 & 90 \\
    0.2 & 21.80 & 90 \\
    0.3 & 30.96 & 90 \\
    0.4 & 38.66 & 90 \\
    0.5 & 45    & 90 \\
    \bottomrule
    \end{tabular}
\end{table}

\begin{figure}[hbt!]
    \centering
    \begin{subfigure}[]{.48\textwidth}
        \centering
        \includegraphics[width=\textwidth]{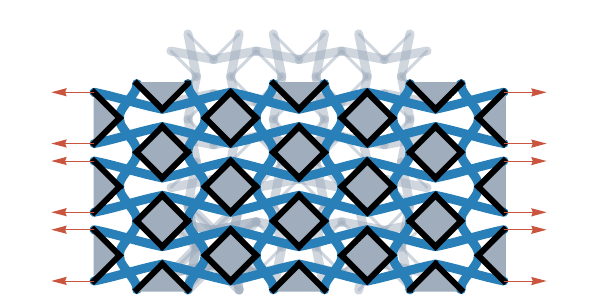}
        \caption{Euler-Bernoulli beam model with $\Lambda=1$, $\phi=0$ yielding $\nu=0.325$}
    \end{subfigure}
    \begin{subfigure}[]{.48\textwidth}
        \centering
        \includegraphics[width=\textwidth]{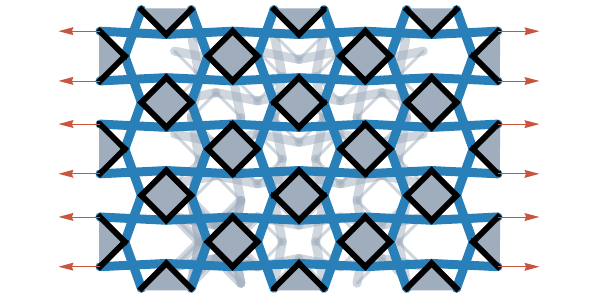}
        \caption{Timoshenko beam model with $\Lambda=1$, $\phi=1.76$ yielding $\nu=-0.2$}
    \end{subfigure}
    \caption{
    As in Fig.~\ref{fig:triangular_elastic_deformed}, except that the lattice is now re-entrant ($\alpha=\pi/4$ and $\eta=0.3$) and includes square rigid inclusions. The Timoshenko model (b) correctly exhibits auxetic behavior, in contrast to the Euler-Bernoulli model (a). 
    }
    \label{fig:re-entrant_sqrigid_lattice_deformed}
\end{figure}

Deformed shapes of the grid with rigid inclusions are shown in Fig.~\ref{fig:re-entrant_sqrigid_lattice_deformed} for both the Euler-Bernoulli model (left) and the Timoshenko model (right), with parameters $\Lambda=1$, $\alpha=\pi/4$, and $\eta=0.3$. The Euler-Bernoulli model incorrectly predicts a positive Poisson's ratio, in agreement with the polar plot in Fig.~\ref{fig:re-entrant_sqrigid_lattice_nu_polar_plot} (left, yellow line). In contrast, the Timoshenko beam grid exhibits lateral expansion along the direction $\be_2$ under horizontal uniform stress, resulting in a negative Poisson's ratio of the equivalent material, as predicted by the yellow line in the right panel of the same figure.

\subsection{An oblique lattice}
\begin{figure}[hbt!]
    \centering
    \begin{subfigure}[]{0.8\textwidth}
        \centering
        \includegraphics[width=\textwidth, keepaspectratio]{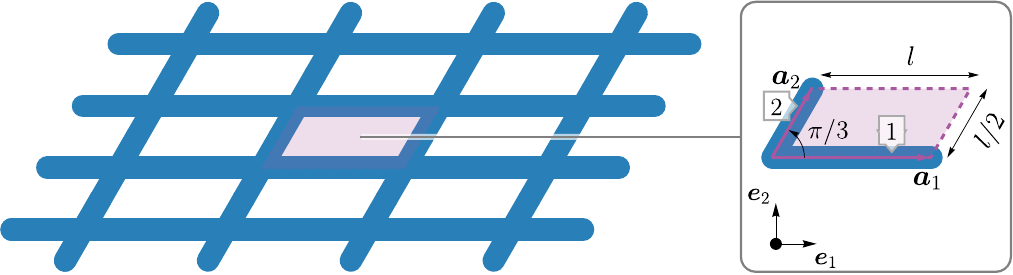}
    \end{subfigure}
    \caption{
    Oblique lattice (left) composed of elastic Timoshenko beams arranged on the plane to form an L-shape pattern inclined at an angle of $\pi/3$ (right).
    }
    \label{fig:oblique_elastic_lattice_geometry}
\end{figure}

An oblique lattice is now analyzed, as illustrated in Fig.~\ref{fig:oblique_elastic_lattice_geometry}. The grid is constructed by tiling a parallelogram-shaped unit cell with side lengths $l$ and $l/2$, oriented along the $\ba_1$ and $\ba_2$ directions, respectively. The $\ba_2$ direction is inclined at an angle of $\pi/3$ with respect to the $\be_1$ axis. The horizontal and oblique beams have rectangular cross-sections with dimensions $b \times h$ and $b/2 \times 2h$, respectively, such that all beams share the same slenderness, $\Lambda = \sqrt{12}\, l/b$, and cross-sectional area, $A = bh$.

\begin{figure}[hbt!]
    \centering
    \begin{subfigure}[]{.48\textwidth}
        \centering
        \includegraphics[width=\textwidth]{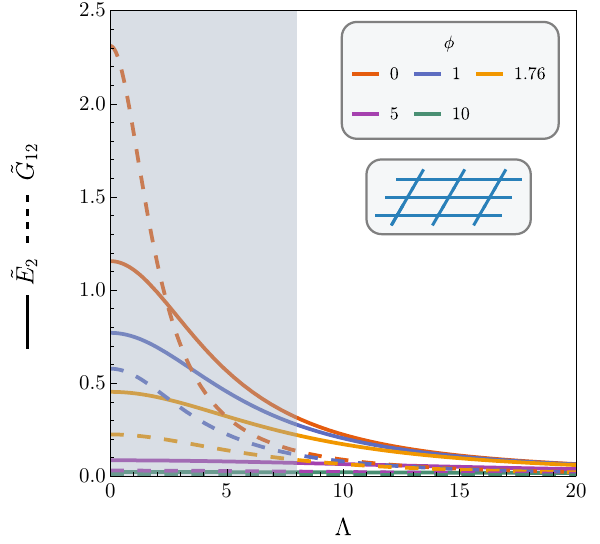}
    \end{subfigure}
    \begin{subfigure}[]{.48\textwidth}
        \centering
        \includegraphics[width=\textwidth]{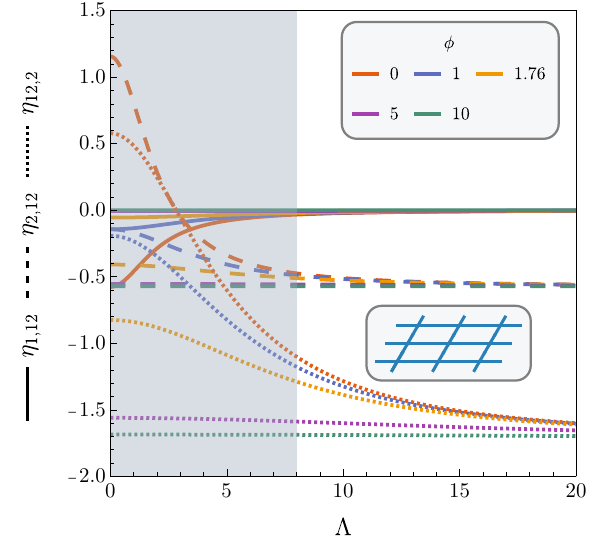}
    \end{subfigure}
    \caption{
    Non-trivial engineering constants of the elastic material equivalent to the oblique grid plotted as functions of the slenderness $\Lambda$, for different values of $\phi$. Left: Dimensionless Young's modulus, $\tilde E_2$, and dimensionless shear modulus, $\tilde G_{12}$; Right:  Coefficients of mutual influence of I and II kind, $\eta_{1,12}$, $\eta_{2,12}$, $\eta_{12,2}$.
    }
\label{fig:oblique_elastic_lattice_E_G_etas-lambda}
\end{figure}

This condition, together with the chosen beam lengths, is sufficient to ensure that the equivalent material is anisotropic, with an elasticity tensor given by
\begin{equation}
\label{eq:oblique_lattice_elasticity_tensor}
    \begin{bmatrix}
        \tilde \fE_{1111} & \tilde\fE_{1122} & \tilde\fE_{1112} \\
        \cdot & \tilde\fE_{2222} & \tilde\fE_{2212} \\
        \cdot & \cdot & \tilde\fE_{1212}
    \end{bmatrix} 
    = \frac{1}{8\sqrt{3}\Theta}
    \begin{bmatrix}
        33\Theta + 24 & 3\Theta - 24 & \sqrt{3}\Theta + 24\sqrt{3}  \\
        \cdot         & 9\Theta + 24 & 3\sqrt{3}\Theta - 24\sqrt{3} \\
        \cdot         & \cdot        & 3\Theta + 72
    \end{bmatrix},
\end{equation}
where dimensionless quantities are defined according to Eq.~\eqref{peretta}, and $\Theta$ depends on both $\Lambda$ and $\phi$, as given by Eq.~\eqref{eq:Theta}. It is worth noting that the elasticity tensor in Eq.~\eqref{eq:oblique_lattice_elasticity_tensor} is fully populated, implying a coupling between axial and shear deformations. This coupling is better understood by examining the engineering constants, which depend on $\Lambda$ and $\phi$ through $\Theta$, and are expressed as
\begin{equation}
    \begin{gathered}
        \tilde E_1 = \frac{4}{\sqrt{3}}, \quad
        \tilde E_2 = \frac{16\,\sqrt{3}}{\Theta+24}, \quad
        \tilde G_{12} = \frac{16}{\sqrt{3}\,(\Theta + 4)}, \\
        \eta_{1,12} = -\frac{4}{\sqrt{3}\,(\Theta + 4)}, \quad
        \eta_{2,12} = -\frac{\Theta - 8}{\sqrt{3}\,(\Theta + 4)}, \quad 
        \eta_{12,1} = -\frac{1}{\sqrt{3}} , \quad
        \eta_{12,2} = -\frac{\sqrt{3}\,(\Theta - 8)}{\Theta + 24},
    \end{gathered}
\end{equation}
while the Poisson's ratios vanish, as also observed for the square lattice discussed in Section~\ref{sec:square_lattice}. The non-trivial engineering constants are plotted in Fig.~\ref{fig:oblique_elastic_lattice_E_G_etas-lambda} as functions of the slenderness $\Lambda$, considering five different values of shear coefficient $\phi$. 

Similarly to the case of the hexagonal grid, bending is essential in the present geometry, as $\tilde E_2$ vanishes when the bending stiffness of the beams approaches zero, enabling rigid-body displacements. This behavior is evident in Fig.~\ref{fig:oblique_elastic_lattice_E_G_etas-lambda} (left), where $\tilde E_2$ tends to zero as $\Lambda$ increases. Although shear deformation softens the equivalent material, reducing the corresponding elastic moduli, this effect decreases as the beams become more slender, and essentially disappears for $\Lambda \approx 20$. Note also that the equivalent shear modulus, $\tilde G_{12}$ is significantly larger for $\phi = 0$ than for $\phi > 0$ (for example, when $\Lambda = 1$, $\tilde G_{12} = 1.85$ for $\phi = 0$, and $\tilde G_{12} = 0.54$ for $\phi = 1$). However, this gap reduces as the beams become slender.

The coefficients of mutual influence, shown on the right of Fig~\ref{fig:oblique_elastic_lattice_E_G_etas-lambda}, strongly depend on both the choice of the beam theory and the slenderness of the beams. For $\Lambda < 2\sqrt{2}$, the coefficients $\eta_{2,12}$ and $\eta_{12,2}$ become positive under the Euler-Bernoulli beam theory, whereas $\eta_{1,12}$ remains negative for all values of $\Lambda$ and approaches zero as the beams become increasingly slender.

\begin{figure}[hbt!]
    \centering
    \begin{subfigure}[]{.48\textwidth}
        \centering
        \includegraphics[width=\textwidth]{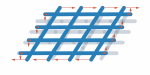}
        \caption{Euler-Bernoulli beam model with $\Lambda=1$, $\phi=0$, yielding $\eta_{12,2} \simeq 0.485$}
    \end{subfigure}
    \begin{subfigure}[]{.48\textwidth}
        \centering
        \includegraphics[width=\textwidth]{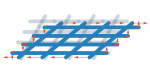}
        \caption{Timoshenko beam model with $\Lambda=1$, $\phi=1.76$, yielding $\eta_{12,2} \simeq -0.841$}
    \end{subfigure}
    \caption{
    A shear stress applied to the oblique grid induces both shear and normal strain components. The Timoshenko model (b) correctly captures the deformation response, while the Euler-Bernoulli model (a) exhibits an incorrect deformation response. 
    }
    \label{fig:oblique_elastic_lattice_deformed}
\end{figure}

Due to anisotropy, the application of a shear stress in the equivalent material induces not only a shear strain but also normal strain components. This behavior is illustrated in Fig.~\ref{fig:oblique_elastic_lattice_deformed}, where a portion of the grid is subjected to shear loading. The case $\phi=0$ (corresponding to the Euler-Bernoulli model) is shown on the left, while $\phi=1.76$ (corresponding to a rectangular cross-section for a Timoshenko beam with $\nu_s=0.3$) is shown on the right. In both cases, homogenization gives $\eta_{12,1} = -1/\sqrt{3}$. However, $\eta_{12,2} \approx 0.485$ for $\phi=0$, leading to vertical dilatation, whereas $\eta_{12,2} \approx -0.841$ for $\phi=1.76$, resulting in vertical contraction.

\section{Conclusions}
\label{sec:concluding}
The Timoshenko beam model provides an accurate description of the mechanical response of periodic grids composed of stubby beams, even at very low slenderness ratios. This accuracy is preserved when either microstructured beams are employed or lattice joints are designed to avoid geometric interference between adjacent beams, for instance, by placing beams on different planes and allowing interaction only through their end cross-sections.

A rigorous equivalent linear elastic continuum can be derived through static homogenization. The results show that stubby beams enable effective mechanical properties in the homogenized material that cannot be realized with slender beams. In addition, the Timoshenko model provides an effective approach for modeling microstructured beams and can yield more reliable predictions than the Euler–Bernoulli model, particularly in relation to auxetic behavior. It also plays an important role in capturing other effective elastic properties.

\section*{Acknowledgements}
All the authors acknowledge financial support from the European Research Council (ERC) under the European Union’s Horizon Europe research and innovation programme (Grant agreement No. ERC-ADG-2021-101052956-BEYOND). The methodologies developed in this work fall within the aims of the GNFM (Gruppo Nazionale per la Fisica Matematica) of the INDAM (Istituto Nazionale di Alta Matematica).

\appendix

\section{Appendix: A finite element validation of the homogenization scheme}
\label{sec:appendix}
\begin{figure}[hbt!]
    \centering
    \begin{subfigure}[]{.60\textwidth}
        \centering
        \includegraphics[width=\textwidth]{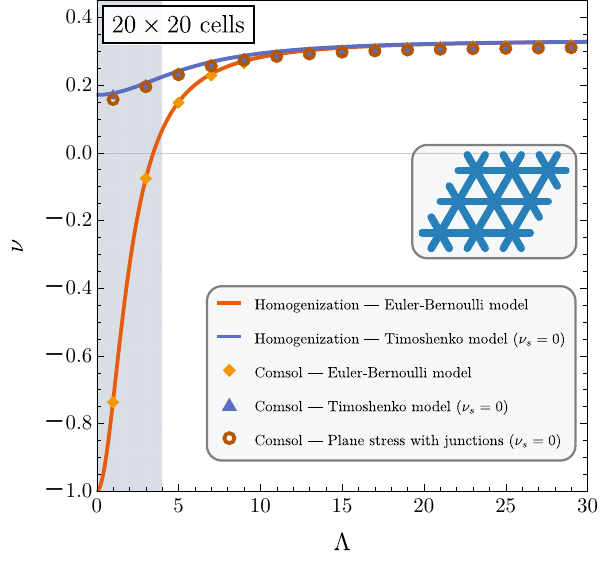}
    \end{subfigure}
    \caption{
    Poisson's ratio of a material equivalent to an equilateral triangular grid made of elastic beams at varying slenderness $\Lambda$. Stubby beams (corresponding to the gray region in the plot) yield Poisson's ratio values unattainable by a solid material with constituent Poisson's ratio $\nu_s =0$. Results are obtained from: (i.) the homogenized response of an infinite lattice using Euler-Bernoulli (solid orange curve) or Timoshenko (solid blue curve) beam theory, Eqs.~\eqref{eq:triangular_lattice_engineering_constants}$_2$ with $\phi=0$ and $\phi=1.55$, respectively; (ii.) finite element analysis of a finite grid ($20 \times 20$ unit cells) of Euler-Bernoulli (yellow diamonds) or Timoshenko (blue triangles) beams; (iii.) finite element analysis of a finite grid ($20 \times 20$ unit cells) discretized with plane stress elements and modeled as separated beams operating in different planes and rigidly jointed at the nodes to avoid interference (red circles), following a scheme similar to that illustrated in Fig.~\ref{fig:decoupled_model}.
    }
    \label{fig:triangular_lattice_poisson_ratio_numerical}
\end{figure}

\begin{figure}[hbt!]
    \centering
    \begin{subfigure}[]{.60\textwidth}
        \centering
        \includegraphics[width=\textwidth]{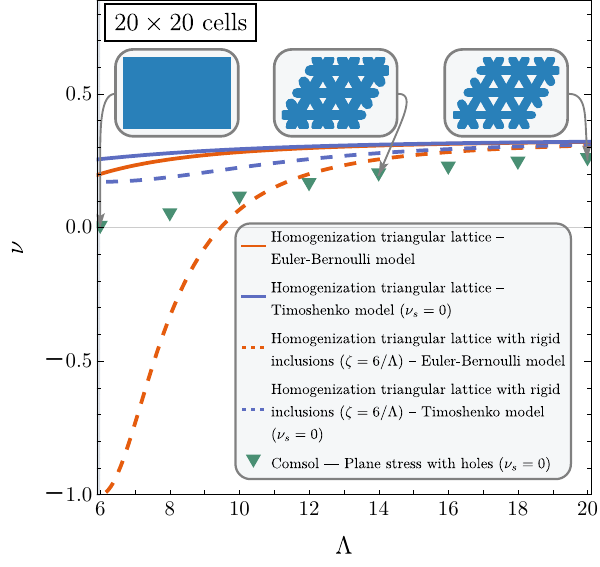}
    \end{subfigure}
    \caption{
    Poisson's ratio for a material equivalent to an equilateral triangular grid of elastic beams, shown as a function of beam slenderness $\Lambda$. 
    Results are obtained from: 
    (i.) the homogenized response of an infinite lattice using Euler-Bernoulli (solid orange curve) and Timoshenko (solid blue curve) beams, as given by Eq.~\eqref{eq:triangular_lattice_engineering_constants}$_2$ with $\phi=0$ and $\phi=1.55$, respectively; 
    (ii.) the same homogenized approach as in (i.), but with rigid inclusions added at the beam junctions (dashed curves), as given by Eq.~\eqref{eq:triangular_hexrigid_equivalent_eng_const}$_2$; 
    (iii.) finite element analysis of a finite grid consisting of $20 \times 20$ unit cells, discretized with plane stress elements and modeled as a solid plate perforated with triangular holes (green triangles). Insets illustrate that the gaps between beams vanish at $\Lambda=6$, corresponding to the limiting case of an intact plate. 
    }
    \label{fig:triangular_lattice_plane_stress_20_cells}
\end{figure}

Numerical simulations, shown in Figs.~\ref{fig:triangular_lattice_poisson_ratio_numerical} and \ref{fig:triangular_lattice_plane_stress_20_cells}, were performed using Comsol Multiphysics 6.2 to validate the homogenization scheme introduced in this work. The results of the homogenization model are also included in the figures as solid lines. In both figures, the effective Poisson's ratio $\nu$ is plotted for an equilateral triangular lattice composed of a material with zero Poisson's ratio ($\nu_s=0$), as a function of the slenderness of the beams connecting the nodes. The slenderness is defined as $\Lambda = l \sqrt{A/J} = 2\sqrt{3}\, l/b$, where $l$ is the distance between the nodes and $b$ is the thickness of the beams.
In addition to the homogenization predictions, numerical results are presented for a finite lattice occupying a domain of $20 \times 20$ cells subjected to uniaxial tension. The effective Poisson's ratio is evaluated based on the transverse contraction measured at points located in the central region of the lattice, in order to reduce the influence of edge effects.  

\begin{figure}[hbt!]
    \centering
    \begin{subfigure}[]{.96\textwidth}
        \centering
        \includegraphics[width=\textwidth]{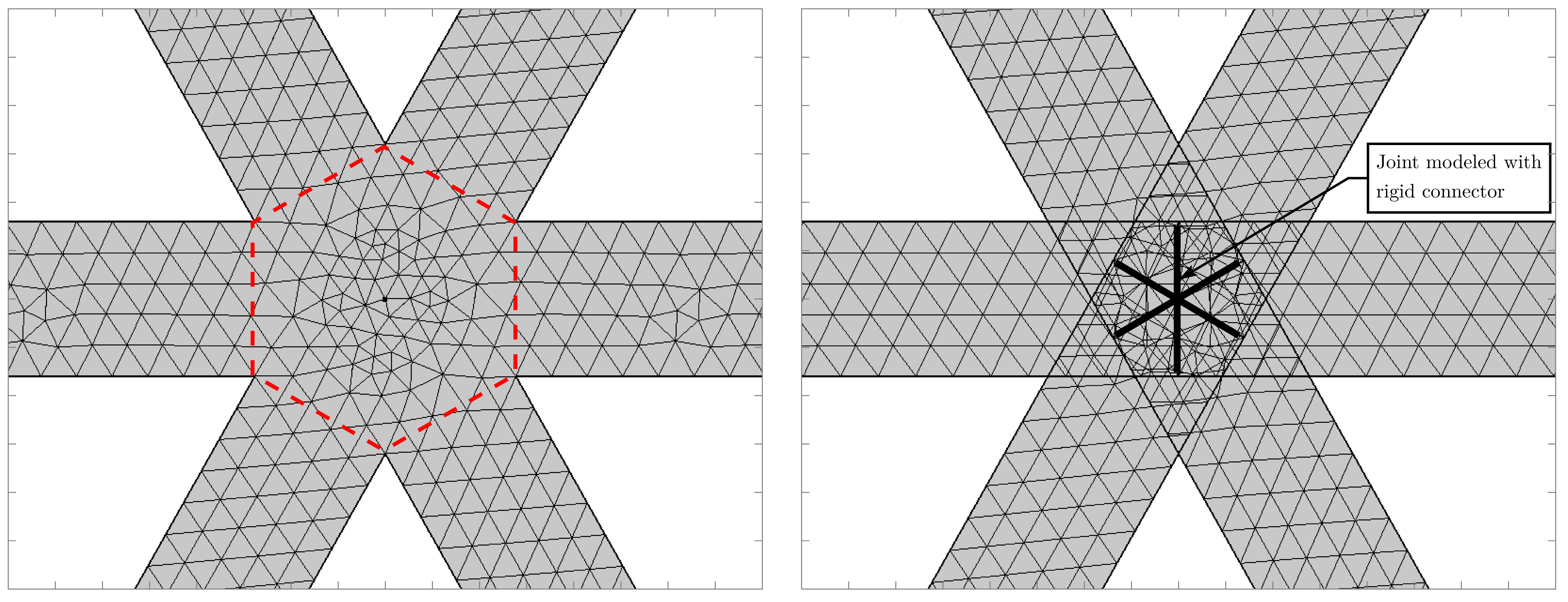}
    \end{subfigure}
    \caption{
    Detail of a node in the equilateral triangular lattice discretized with plane stress elements. The geometry is modeled in two configurations: as a solid plate with triangular holes (left) and as three separate beams lying on different planes and connected through a rigid joint (right). In the latter case, each beam is defined as an independent physics domain and meshed individually. The rigid joint is implemented using Euler-Bernoulli beam elements, made rigid by enabling the {\em rigid connector} feature available in Comsol.
    }
    \label{fig:plane_stress_meshes}
\end{figure}

In the lattice analyzed for Fig.~\ref{fig:triangular_lattice_poisson_ratio_numerical}, the beams are modeled in three different ways: 
\begin{itemize}
    \item (i.) Using beam elements based on the Euler-Bernoulli theory;
    \item (ii.) Using beam elements based on the Timoshenko theory;
    \item (iii.)  
    Using plane stress elements, with joints modeled according to the conceptual prototype shown in Fig.~\ref{fig:decoupled_model}, to avoid geometric interference between beams converging at the nodes. Results from this model are labeled `Plane stress with junctions' in the figure. In this configuration, the beams lie on different planes and interact only through their end cross-sections, which are connected by rigid connectors, see the right panel in Fig.~\ref{fig:plane_stress_meshes}. This layered arrangement allows beam overlap by eliminating geometric interference at the nodes.
    In the Comsol finite element simulation, each beam is defined as an independent physics domain and connected to others through rigid joints, represented by thick lines in the right panel of Fig.~\ref{fig:plane_stress_meshes}. These rigid joints are implemented using Euler-Bernoulli beam elements endowed with the {\em rigid connector} property enabled.
\end{itemize}

Figure \ref{fig:triangular_lattice_poisson_ratio_numerical} shows that the homogenization scheme performs remarkably well, with all numerical results closely matching the continuous curves obtained from the homogenization model. In particular, the simulations in which the beams connecting the nodes are implemented using plane stress elements, the above case (iii.), coincide with those based on the Timoshenko theory and progressively deviate from the Euler-Bernoulli approximation as the slenderness $\Lambda$ decreases. 
This observation further emphasizes the importance of accounting for shear deformability in the beam model, especially when dealing with low slenderness ratios.

The composite corresponding to the results plotted with green triangles in Fig.~\ref{fig:triangular_lattice_plane_stress_20_cells} refers to a plate under plane stress, perforated with triangular holes arranged to obtain a triangular lattice. These results are labeled \lq Plane stress with holes' in the figure. The beams forming the perforated plate are monolithically connected at the nodes, as shown in the left panel of Fig.~\ref{fig:plane_stress_meshes} and in the insets of Fig.~\ref{fig:triangular_lattice_plane_stress_20_cells}. It can be anticipated that, as the holes in the plate become increasingly small and eventually disappear, a beam-based homogenization scheme loses validity. In such cases, a more appropriate homogenization approach (not pursued here) would involve modeling the structure as a plate with a periodic distribution of voids. Although a beam-based approximation of the perforated plate is clearly inadequate at low slenderness, the results are nevertheless compared with four alternative beam-based models as follows: 
\begin{itemize}
    \item The orange (Euler-Bernoulli) and blue (Timoshenko) {\it continuous} lines correspond to the homogenization scheme introduced in this article, where the effective Poisson's ratio is given by Eq.~\eqref{eq:triangular_lattice_engineering_constants}$_2$. 
    \item The orange (Euler-Bernoulli) and blue (Timoshenko) {\it dashed} lines also refer to homogenization, but in this case, the hexagonal connecting region at the nodes (marked by red dashed lines in the left panel of Fig.~\ref{fig:plane_stress_meshes}) is treated as a rigid inclusion. The perforated plate is thus modeled as a triangular beam lattice with rigid inclusions (see section \ref{rigid_inclusions} and Fig.~\ref{fig:triangular_hexrigid_lattice_geometry}). 
    In particular, the effective Poisson's ratio is computed from Eq.~\eqref{eq:triangular_hexrigid_equivalent_eng_const}$_2$, where the slenderness $\Lambda$ is rescaled as $\Lambda(1-\zeta)$ to ensure consistent comparison with the triangular lattice without inclusions. The parameter $\zeta$, which characterizes the size of the rigid inclusion, is set to $6/\Lambda$, corresponding to the perforated plate configuration.
\end{itemize}    

The insets in Fig.~\ref{fig:triangular_lattice_plane_stress_20_cells} are drawn to scale and illustrate how a plate with triangular holes gradually {\em condenses} into an intact plate as the elements connecting the holes become sufficiently stubby. In particular, the intact plate is obtained when the slenderness reaches $\Lambda=6$. 
For a constituent material having $\nu_s=0$, the effective Poisson's ratio $\nu$ of the material equivalent to the perforated plate varies between approximately $1/3$ and $0$, with $\nu = 0$ corresponding to the intact plate. The figure shows that all models consistently tend to $\nu=1/3$ at high slenderness. However, at low slenderness, only the model using Timoshenko rods with rigid inclusions exhibits the correct trend, followed by the results for the perforated plate. Anyway, the quantitative values for the equivalent $\nu$ remain visibly different. 

\printbibliography

\end{document}